\documentclass[twocolumn,dvipsnames]{aastex62}
\pdfoutput=1 
\usepackage{amsmath,amstext}
\usepackage{apjfonts}
\usepackage[figure,figure*]{hypcap}
\usepackage{ulem}
\usepackage{xspace}
\graphicspath{ {./figures/} }
\DeclareGraphicsExtensions{.png,.pdf}
\newcolumntype{H}{>{\setbox0=\hbox\bgroup}c<{\egroup}@{}}
\usepackage{lineno}
\usepackage{outlines}
\usepackage{xfrac}
\usepackage{graphicx}
\usepackage{natbib}
\usepackage{amssymb}
\usepackage{amsmath}
\usepackage{array,booktabs}
\usepackage{mathptmx}
\usepackage{booktabs}
\usepackage{float}
\usepackage{todonotes}
\usepackage{wrapfig}

\usepackage{multirow}

\newcommand{\mej}{\,{M_{\rm ej}}}
\newcommand{\s}{\,{{\rm s}}}
\newcommand{\msun}{\,{M_\odot}}
\newcommand{\mbh}{\,{M_{\rm BH}}}
\newcommand{\abh}{\,{a_{\rm BH}}}
\newcommand{\rg}{\,{r_{\rm g}}}
\newcommand{\ye}{\,{Y_{\rm e}}}

\newcommand{\mycomment}[1]{}
\newcolumntype{P}[1]{>{\centering\arraybackslash}p{#1}}

\newcommand{\pBw}{pBw}
\newcommand{\pBs}{pBs}
\newcommand{\cBs}{cBs}
\newcommand{\cBw}{cBw}
\newcommand{\cBwz}{cBwA0}
\newcommand{\pBz}{pB0}
\newcommand{\hammer}{\texttt{H-AMR}}
\DeclareSymbolFont{cmletters}{OML}{cmm}{m}{it}
\DeclareMathSymbol{v}{\mathalpha}{cmletters}{"76}

\shorttitle{Magnetically-Driven Neutron-Rich Ejecta Unleashed}
\shortauthors{Issa et al.}

\begin{document}

\title[Magnetically-Driven Neutron-Rich Ejecta Unleashed]{Magnetically-Driven Neutron-Rich Ejecta Unleashed: Global 3D Neutrino-General Relativistic Magnetohydrodynamic Simulations of Collapsars Probe the Conditions for $r$-process Nucleosynthesis}

    \author[0009-0005-2478-7631]{Danat Issa}
    \email{danat@u.northwestern.edu}
	\affiliation{Center for Interdisciplinary Exploration \& Research in Astrophysics (CIERA), Physics \& Astronomy, Northwestern University, Evanston, IL 60201, USA}

    \author[0000-0003-3115-2456]{Ore Gottlieb}
    \affiliation{Center for Computational Astrophysics, Flatiron Institute, New York, NY 10010, USA}
    \affiliation{Department of Physics and Columbia Astrophysics Laboratory, Columbia University, Pupin Hall, New York, NY 10027, USA}
    
    \author[0000-0002-3635-5677]{Brian D. Metzger}
    \affiliation{Department of Physics and Columbia Astrophysics Laboratory, Columbia University, Pupin Hall, New York, NY 10027, USA}
    \affiliation{Center for Computational Astrophysics, Flatiron Institute, New York, NY 10010, USA}
    
    \author[0000-0003-2982-0005]{Jonatan Jacquemin-Ide}
    \affiliation{Center for Interdisciplinary Exploration \& Research in Astrophysics (CIERA), Physics \& Astronomy, Northwestern University, Evanston, IL 60201, USA}
    
    \author[0000-0003-4475-9345]{Matthew Liska}
    \affiliation{Center for Relativistic Astrophysics, Georgia Institute of Technology, Howey Physics Bldg, 837 State St NW, Atlanta, GA, 30332, USA}
    \affiliation{Institute for Theory and Computation, Harvard University, 60 Garden Street, Cambridge, MA 02138, USA}
    
    \author[0000-0003-4617-4738]{Francois Foucart}
    \affiliation{Department of Physics \& Astronomy, University of New Hampshire, 9 Library Way, Durham, NH 03824, USA}

    \author[0000-0002-7232-101X]{Goni Halevi}
    \affiliation{Department of Physics, Illinois Institute of Technology, Chicago, IL 60616, USA}
    \affiliation{Center for Interdisciplinary Exploration \& Research in Astrophysics (CIERA), Physics \& Astronomy, Northwestern University, Evanston, IL 60201, USA}

    \author[0000-0002-9182-2047]{Alexander Tchekhovskoy}
    \affiliation{Center for Interdisciplinary Exploration \& Research in Astrophysics (CIERA), Physics \& Astronomy, Northwestern University, Evanston, IL 60201, USA}

\begin{abstract}
Collapsars -- rapidly rotating stellar cores that form black holes -- can power gamma-ray bursts (GRBs) and are proposed to be key contributors to the production of heavy elements in the Universe via the rapid neutron capture process ($r$-process). Previous neutrino-transport collapsar simulations have been unable to unbind neutron-rich material from the disk. However, these simulations have not included sufficiently strong magnetic fields and the black hole (BH), both of which are essential for launching mass outflows. We present \textsc{$\nu$h-amr}, a novel neutrino-transport general relativistic magnetohydrodynamic ($\nu$GRMHD) code, which we use to perform the first 3D global $\nu$GRMHD collapsar simulations. We find a self-consistent formation of a weakly magnetized dense accretion disk, which has sufficient time to neutronize. Eventually, substantial magnetic flux accumulates near the BH, becomes dynamically important, leads to a magnetically arrested disk (MAD), and unbinds some of the neutron-rich material. However, the strong flux also hinders accretion, lowers density, and increases neutrino cooling timescale, which prevents further disk neutronization. Typical collapsar progenitors with mass accretion rates, $\dot{M} \sim 0.1-1 M_\odot/\rm{s}$, do not produce significant neutron-rich ($Y_\text{e} < 0.25$) ejecta. However, we find that MADs at higher mass accretion rates, $\dot{M} \gtrsim \text{few}\, M_\odot/\rm{s}$ (e.g., for more centrally concentrated progenitors), can unbind $M_\text{ej}\lesssim{}M_\odot$ of neutron-rich ejecta. The outflows inflate a shocked cocoon that mixes with the infalling neutron-poor stellar gas and raises the final outflow $Y_\text{e}$; however, the final $r$-process yield may be determined earlier at the point of neutron capture freeze-out. Future work will explore under what conditions more typical collapsar engines become $r$-process factories.

\end{abstract}

\section{Introduction}\label{sec:introduction}

One of the fundamental questions in nuclear astrophysics is the origin of heavy elements ($A\geq 69$), approximately half of which are formed via the rapid neutron capture ($r$-process) nucleosynthesis. The proposed astrophysical sites that generate these heavy nuclei can be divided into two broad classes, both requiring similar conditions in their central engines \citep[see][for reviews]{Metzger2019,Arnould2020,Cowan2021,SiegelReview2022}: (i) black hole (BH)--neutron star (NS) and NS--NS mergers, and (ii) the collapse of massive stellar cores. The former site was confirmed through optical observations of the NS--NS merger GW170817 \citep[see, e.g.,][for reviews]{Nakar2020,Margutti2021}, where a neutron-rich outflow produced heavy nuclei \citep{Kasen2017} whose decay powered the kilonova emission \citep{Metzger2010}.

Generally, the $r$-process nucleosynthesis channel requires neutron-rich outflows. Accretion disks that form during the collapse of rapidly rotating massive stars exhibit similar density and temperature conditions as those found in postmerger disks. Moreover, the larger size and lifetime of collapsar disks render them more massive than postmerger disks, suggesting their potential as the major source of the Universe's heavy $r$-process elements \citep{siegel_collapsars_2019}. However, the stellar infall, which forms collapsar disks, likely obstructs the ejection of neutron-rich disk outflows. Describing this complex, nonlinear behavior necessitates numerical simulations to accurately model the formation of the collapsar disk, the synthesis of heavy elements, and the propagation of outflows through the collapsing star.

The collapse of a massive stellar core initially forms a proto-NS. The $r$-process nucleosynthesis can happen in NS outflows, which are a combination of neutrino-driven \citep[e.g.,][]{Thompson+04,Metzger2007} and magnetically driven \citep{Kotake2004,Ardeljan2005,Moiseenko2006,Obergaulinger2006,Obergaulinger2020,Obergaulinger2021,Mosta2014,Kuroda2020,Aloy2021}. Neutrino-heated winds can facilitate favorable conditions for the $r$-process only for rapidly rotating, strongly magnetized proto-NSs, thanks to the shorter expansion timescales \citep{Thompson+04,Metzger2007,Metzger2008a,Vlasov2014,Vlasov2017,Desai2022,Prasanna2023}. Studies of magnetorotational supernovae (MR-SNe) show that they require rapid rotation and strong magnetic fields to launch powerful magnetically driven outflows, or jets \citep{Nishimura2006,Winteler2012,Nishimura2015,Tsujimoto2015,Halevi2018,Mosta2018,Reichert2021,Reichert2023}. Strong outflows help to unbind neutron-rich matter and drive it at high velocities to avoid neutrino irradiation from the proto-NS, leading to robust $r$-process nucleosynthesis. Weaker initial magnetic field strengths lead to weaker outflows or delayed jet launching and therefore result in the $r$-process element production only up to the second peak \citep{Nishimura2006,Winteler2012,Nishimura2015}. Moreover, effects like current-driven kink instabilities and misalignment between the magnetic and rotation axes can also reduce the ability of MR-SNe to contribute to heavy $r$-process enrichment \citep{Halevi2018,Mosta2018}.

Alternatively, neutron-rich accretion disks can also form around BHs in collapsars \citep[][]{Woosley1993,MacFadyen1999,MacFadyen2001,Fujimoto2004,Pruet2004,Kohri2005,Surman2006,Woosley2006} or in common-envelope jetted explosions \citep{Grichener2019}. Growing computational power and advances in computational methods have enabled the numerical simulations of the $r$-process nucleosynthesis in this context. Earlier calculations showed neutron-rich material production and successful $r$-process operation in collapsar jets \citep{Fujimoto2007,Ono2012,Ko2013}. More recently, simulations modeled the magnetized evolution of an idealized isolated preset torus in hydrostatic equilibrium, i.e., without including the progenitor star \citep{siegel_collapsars_2019,miller_full_2020}. These works show that collapsar disks may become neutron rich. However, it remains unclear whether the behavior of these idealized tori resembles that of disks self-consistently formed during the stellar collapse and whether the neutron-rich elements can overcome the ram pressure of the infalling stellar gas and escape from the collapsing star. More recently, simulations included the collapsing star for the first time, which hindered the ejection of the heavy $r$-process elements \citep{Just2022,fujibayashi_collapse_2023,Dean2024,Dean2024b}. However, these simulations lacked important physics: they reduced the dimensionality to 2D and excluded magnetic effects. The absence of magnetic fields does not allow magnetized outflows to form via, e.g., the Blandford-Znajek \citep[BZ;][]{Blandford1977} and Blandford-Payne--like \citep{blandford_hydromagnetic_1982} processes, which may potentially carry away some of the neutron-rich elements.

In this Letter, we present the first 3D neutrino-transport general-relativistic magnetohydrodynamic ($\nu$GRMHD) collapsar simulations. Our simulations use a two-moment (M1) neutrino-transport scheme. 
We find that including both the self-consistent disk formation in the collapsing stellar gas and subsequent magnetic launching of collimated disk outflows is crucial for the development of neutron-rich ejecta in collapsars. 
We begin by discussing the theoretical requirements for forming and ejecting neutron-rich material in~\S\ref{sec:theory}. Using these arguments, we motivate our numerical setup in~\S\ref{sec:setup}. We present our simulation results in~\S\ref{sec:results} and discuss their implications in~\S\ref{sec:summary}.

\section{How to Make $r$-process Ejecta}\label{sec:theory}

Here, we outline the conditions under which collapsar disks form and generate neutron-rich material, and subsequently eject it. An accretion disk forms if the collapsing stellar gas possesses specific angular momentum, $ l(r) $, exceeding that at the innermost stable circular orbit (ISCO), $ r_{\rm ISCO} $:
\begin{equation}
    l (r) > \sqrt{G\mbh r_{\rm ISCO}}\,,
\end{equation}
where $ r $ is the spherical polar radius and $ \mbh $ is the BH mass, and for simplicity, we use a nonrelativistic approximation for the specific angular momentum at the ISCO. There are theoretical indications that collapsar progenitors feature rapid rotation, which results in early disk formation \citep[e.g.,][]{Gottlieb2024}.

A necessary, but not sufficient, condition for the $r$-process is that the disk must become neutron rich. The degree of neutron richness is inversely correlated with the electron fraction $\ye=n_{\rm p} / (n_{\rm p}+n_{\rm n})$, where $n_{\rm p}$ and $n_{\rm n}$ are the proton and neutron number densities, respectively. For weak interactions to drive the disk material neutron-rich ($\ye<0.5$), it must become dense and mildly degenerate through neutrino cooling (e.g., \citealt{Beloborodov03}). Specifically, when the viscous heating rate is counterbalanced by neutrino cooling, the disk transitions from a thick advective state to an efficiently neutrino-cooled regime (e.g., \citealt{DiMatteo+02}), in which it can become notably dense and neutron rich. Satisfying this ``ignition'' condition requires a minimum mass accretion rate at the ISCO \citep{Chen2007,Metzger2008b,Metzger2008a},
\begin{equation}\label{eq:Mign}
    \dot{M}_{\rm ign} = K_{\rm ign}\,\left(\frac{\alpha_{\rm{eff}}}{0.1}\right)^{5/3}\left(\frac{M_{\rm BH}}{3M_{\odot}}\right)^{4/3}\,\msun\,\s^{-1}\,,
\end{equation}
where $\alpha_{\rm eff}$ is the effective viscosity and $K_{\rm ign}$ is a numerical prefactor that depends on the BH spin $a_{\rm BH}$: $K_{\rm ign}=0.07$ for $a_{\rm BH}=0$ and $K_{\rm ign}=0.02$ for $a_{\rm BH}=0.95$.

If $\dot{M} \gtrsim \dot{M}_{\rm ign}$, there exists an ignition radius, $R_{\rm ign} \gtrsim r_{\rm ISCO}$, within which the disk efficiently cools through electron-positron pair-capture processes:
\begin{equation}
\begin{split}
    p + e^{-} &\rightarrow n + \nu_{\rm e}\,, \\
    n + e^{+} &\rightarrow p + \bar{\nu}_{\rm e}\,,
\end{split}
\end{equation}
and consequently becomes neutron rich. At the ignition radius, $R = R_{\rm ign}$, the neutrino cooling and viscous heating balance each other. In classical accretion disk theory (e.g., for a standard \citealt{shakura_black_1973} disk), a single viscosity coefficient, $\alpha_{\rm eff}$, governs both the viscous heating and accretion. However, in the presence of strong magnetic fields, large-scale magnetic torques can render the accretion more efficient and decouple it from the heating \citep{blandford_hydromagnetic_1982,ferreira_magnetized_1995}. In practice, for strongly magnetized disks, the accretion timescale is a factor of a few shorter than the viscous heating timescale \citep{manikantan_winds_2024}. Therefore, we can define the ignition radius as the radius where the neutrino-cooling timescale, $ t_{\rm cool} $, matches the accretion timescale, $ t_{\rm acc} $:
\begin{equation}\label{eq:Rign}
    t_{\rm cool}(R_{\rm ign})\simeq t_{\rm acc}(R_{\rm ign})\,.
\end{equation}
Here, the cooling timescale is \citep{Metzger2008a}
\begin{equation}\label{eq:tcool}
    t_{\rm cool} \sim \frac{\langle N_{p} \rangle_\rho^{\theta,\varphi}}{\langle \dot{N}_{e^- p} \rangle_\rho^{\theta,\varphi}}\,,
\end{equation}
and the accretion timescale is
\begin{equation}\label{eq:tacc}
    t_{\rm acc} \sim \frac{r}{\langle v^{\hat{r}} \rangle_\rho^{\theta,\varphi}}\,,
\end{equation}
where $\dot{N}_{e^- p}$ is the rate of electron capture on protons and $\langle\cdot\rangle^{\theta,\varphi}_{\rho}$ is a density-weighted average over angles $\theta,\varphi$, which for quantity $X$ is
\begin{equation} \label{eq:avg_def}
    \langle X \rangle_{\rho}^{\theta,\varphi} (r,t) = \frac{\int \sqrt{-g}\rho X {\rm sin}\theta d\theta d\varphi}{\int \sqrt{-g}\rho {\rm sin}\theta d\theta d\varphi}
\end{equation}
where $\sqrt{-g}$ is the metric determinant. We will justify this definition of $ R_{\rm ign} $ in \S\ref{sec:results}.

Whereas disk winds may eject neutron-rich material in compact binary mergers \citep[e.g.,][]{Fernandez2013,Siegel2017,Radice2018}, they might be too weak to unbind the neutron-rich material in the inner parts of collapsar disks \citep[e.g.,][]{Dean2024}. However, powerful relativistic BZ jets, associated with gamma-ray bursts (GRBs), can explode the star and eject the neutron-rich material out of the collapsing star. The formation of such jets becomes possible once enough poloidal (i.e., pointing in the $r$- and $\theta$-directions) magnetic flux accumulates on the BH to overcome the ram pressure of the infalling gas. Namely, when the Alfv\'{e}n velocity becomes comparable to the free-fall velocity, the jet can emerge from the BH event horizon \citep{Komissarov2009,Gottlieb2023}.

The jet-launching criterion is determined by the dimensionless magnetic flux on the BH, defined as
\begin{equation}\label{eq:phiBH}
    \phi \equiv \frac{\Phi}{\sqrt{\dot{M}\rg^2 c}} \sim \frac{B \rg^2}{\sqrt{\rho v_r \rg^4 c}}\,,
\end{equation}
where $ \rg \equiv GM_{\rm BH} / c^2 $ is the BH gravitational radius. To infer the minimum value of $\phi$ for jet launching, we use estimates by \cite{Komissarov2009}, where they found $\phi_\mathrm{crit} \sim 20$ in idealized simulations of jet launching for radial magnetic geometry (this corresponds to $\kappa_{\rm c}\sim1.5$ in their work). In addition to powerful BH-powered outflows, magnetically arrested disks \citep[MADs;][]{Narayan2003,Tchekhovskoy2011} can also launch massive disk winds whose properties we study below. The accretion flow reaches the maximum magnetization when it goes to MAD, which occurs at $ \phi \approx 50 $ \citep{Tchekhovskoy2011}.

We conclude that collapsars can eject $ r $-process elements if (1)~they form an accretion disk with high mass accretion rates and efficient neutrino cooling, and (2)~once the $ r $-process elements are synthesized in the disk, dynamically important magnetic fields around the BH launch magnetized jets and winds that manage to unbind the neutron-rich material.

\section{Numerical setup and method}\label{sec:setup}

\subsection{Initial conditions}\label{sec:grmhd}

\begin{table*}
	\centering
	\caption{Collapsar model parameters. The leftmost column: model name, which follows the [cp]B[0ws][A\#] convention, where c or p indicates stellar density profile (core or power law, respectively); 0, w, or s indicates the magnetic field strength (zero, weak, and strong field); and A\# indicates BH spin (omitted for the fiducial value, $a_{\rm BH}=0.8$). The rest of the columns, from left to right:  initial density slope, $\alpha_{\rm p}$; grid resolution, $N_r\times N_\theta\times N_\phi$; simulation duration, $t_{\rm f}$; maximum initial core magnetic field strength, $B_{\rm core}$; MAD onset time, $t_{\rm MAD}$; jet (outflow) launch time, $t_{\rm jet}$; $\dot M$ at these two times; jet power at the larger of these two times, $L_{\rm j}$; and total mass of neutron-rich ($\ye<0.25$) material at $t_{\rm MAD}$, $M_{\rm nr}$. Parameter values common to all models are BH mass, $M_{\rm BH}=4\msun$; stellar envelope total mass $M_{\rm star}=70\msun$; stellar radius, $R_{\rm star}=4\times 10^{10}$~cm; and magnetic core radius, $R_{\rm core}=10^8$~cm (such that $\mu=B_{\rm core} R_{\rm core}^2/2$ in Eq.~\ref{eq:Aphi}).}
	\begin{tabular} {|c|@{\,}P{25mm}|c|P{17mm}|P{11mm}|@{\,}P{7mm}@{\,}|@{\,}P{7mm}@{\,}|@{\,}P{12mm}@{\,}|@{\,}P{9mm}@{\,}|@{\,}P{11mm}@{\,}|@{\,}P{14mm}@{\,}|} 
		\hline
		Model & $\rho\propto r^{-\alpha_{\rm{p}}}$ & $N_r\times N_\theta \times N_\phi$ & $t_{\rm f}$ [$10^3\rg/c$] and [$\text s$] & $B_{\rm core}$ [$10^{12}$ G] & $t_{\rm MAD}$ [s] & $t_{\rm jet}$ [s] & $\dot{M}(t_{\rm MAD})$ [$\msun$/s]& $\dot{M}(t_{\rm jet})$ [$\msun$/s] & $L_{\rm j}$ [$10^{53}$ erg/s] & $M_{\rm nr}(t_{\rm MAD})$ [$\msun$]\\
		\hline
        \pBw & \multirow{3}{*}{\hfil $\alpha_{\rm{p}}=1.5$} & $336\times384\times64$ & $51.3$k\,=\,$1.01$\,s & $1.9$ & -- & -- & -- & -- & -- & --\\
        \pBs & & $192\times128\times32$ & $49.6$k\,=\,$0.98$\,s & $5.5$ & 0.17 & 0.06 & 0.1 & $0.3$ & $1.1$ & 0 \\
        \pBz & & $192\times128\times64$ & $35.1$k\,=\,$0.71$\,s & -- & -- & -- & -- & -- & -- & --\\
        \hline
		\cBs & \multirow{3}{*}
  {\hfill 
        \begin{minipage}[b][1cm][c]{3cm}
        \vspace{-0.2cm}                 
        \begin{equation*}
          \alpha_{\rm{p}} =
            \begin{cases}
              0,\phantom{.5} r<R_{\rm core} &\\
              2.5, r>R_{\rm core} &
            \end{cases}       
        \end{equation*}
        \end{minipage}
    }        
        & $192\times128\times64$ & $84.2$k\,=\,$1.66$\,s & $71$ & 0.04 & 0.02 & 5.9 & $3.0$ & $44$ & 0.25 \\
        \cBw & & $288\times288\times64$ & $51.6$k\,=\,$1.02$\,s & $28$ & 0.23 & 0.08 & 2.8 & $16.4$ & $23$ & 0.94 \\		
        \cBwz & & $288\times288\times64$ & $49.7$k\,=\,$0.98$\,s & $28$ & 0.33 & 0.40 & 3.6 & $3.0$ & $0.21$ & 0.31 \\
		\hline	
	\end{tabular}
     \label{tab:models}
\end{table*}

To determine the electron fraction evolution, we implement a novel two-moment (M1) neutrino-transport scheme alongside the Helmholtz equation of state (EOS; see \S\ref{sec:nu_transport}) into the GPU-accelerated code \textsc{h-amr} \citep{hamr_paper}. Using the resulting \textsc{$\nu$h-amr} code, we conduct a suite of 3D $\nu$GRMHD simulations of collapsing stars, as outlined in Table~\ref{tab:models}. In the following, we adopt the units $G=c=1$ and use Heaviside-Lorentz units for magnetic fields, thereby absorbing the factor of $1/\sqrt{4\pi}$ into the definition of magnetic field strength, $B$.

We set up our collapsar simulations following \citet{gottlieb_black_2022,gottlieb_black_2022-1}. We set the newly formed BH mass to $\mbh =4\,\msun$ and its dimensionless spin magnitude to $\abh=0.8$. Our simulation incorporates a static metric such that the BH mass and spin do not evolve over the course of the simulation. The BH is embedded in a stripped-envelope (e.g., Wolf--Rayet-like) star of mass\footnote{As we will show, at the relevant simulation times (MAD onset timescale), the accreted mass is $ \lesssim 10\,\msun $. This implies that a slightly more extended core with a steeper density profile outside will produce the same accretion rates with $ M_\star \approx 20\,\msun $.} $M_{\star}=70\,\msun$ and radius $R_{\rm star}=4\times 10^{10}\,\rm{cm}$ \citep{WoosleyHeger2006}. We adopt a (broken) power law as the radial stellar mass density profile:
\begin{equation}
	\rho(r) \propto r^{-\alpha_{\rm p}} \left( 1 - \frac{r}{R_{\rm star}} \right)^3\,,
\end{equation}
where the density power-law indices $\alpha_{\rm p}$ of our simulations are listed in Table~\ref{tab:models}. We simulate two progenitor models: a typical collapsar with $\alpha_{\rm p} = 1.5$, representing the density profile established in the aftermath of BH formation \citep{halevi_density_2023}; and a collapsar with an extreme, more centrally concentrated distribution with a constant density core of radius $10^8,\rm{cm}$, followed by a density slope of $\alpha_{\rm p} = 2.5$, to probe neutron-rich outflow ejection in the regime of high mass accretion rates.

The initial thermal pressure profile in the star is described by
\begin{equation}
	P_{\rm gas}(r) = P_{0} \frac{\rho(r)}{r / \rg}\,.
\end{equation}
where $P_{0} = 0.1$ for models without and $P_{0} = 0.25$ for models with constant density core. This prefactor is chosen such that the pressure on the entire grid is within the temperature validity range of the EOS (\S~\ref{sec:eos}). The initial thermal pressure is unlikely to impact the disk dynamics since the pressure in the disk will be dominated by energy dissipation due to turbulence associated with the accretion process. Consequently, the primary effect of the initial gas pressure is to provide additional support against gravity, potentially delaying the disk formation and the subsequent infall of stellar gas onto the accretion disk.

At the onset of the collapse, the gas velocity is purely azimuthal. We choose the radial rotation profile such that each spherical shell rotates at a constant angular frequency whose radial profile we express via the specific angular momentum:
\begin{equation}
	l(r, \theta) = 
	\begin{cases}
		\left( \displaystyle \frac{r}{r_{\rm rot}} \right)^2 \sqrt{\mbh r_{\rm circ}} \sin^2\theta, &\quad r < r_{\rm rot}\, \\
		\sqrt{\mbh r_{\rm circ}} \sin^2\theta, &\quad r > r_{\rm rot}\,,
	\end{cases} 
\end{equation}
where $r_{\rm circ} = 25 \rg$ is the circularization radius, and we choose the radius of rigid rotation, $r_{\rm rot} = 70\rg$, such that the flow is sub-Keplerian inside $r_{\rm rot}$. This facilitates disk formation once the shell at this radius reaches the BH horizon on the timescale $t_{\rm f-f} \approx r_{\rm rot}^{1.5}/\sqrt{2}\approx 8 \rm{ms}$. This is consistent with the expectations for collapsar progenitors \citep{Gottlieb2024}. The equatorial inflow faces a centrifugal barrier at $r = r_{\rm circ}$, whereas polar inflows fall onto the BH.

We adopt an initial magnetic field described by the covariant vector potential:
\begin{equation}\label{eq:Aphi}
	A_{\varphi} (r, \theta) = \mu \sin^2\theta \cdot \max \left[ \frac{r^2}{r^2 + R_{\rm core}^2} - \left( \frac{r}{R_{\rm star}} \right)^3, 0 \right]\,,
\end{equation}
where we adopt a stellar core radius, $R_{\rm core} = 10^8 \,\rm{cm}$. This configuration results in a nearly uniform vertical magnetic field at $r \lesssim R_{\rm core}$ that turns radial inside the star and closes near the stellar surface. The magnetic moment $\mu$ is chosen so that magnetization in the core is $\sigma={b^2}/{\rho}\lesssim 0.1$, corresponding to the gas-to-magnetic-pressure ratio of $\beta = \max P_{\rm gas} / \max P_{\rm mag} \sim$ few.

We use a uniform computational grid in $\log_{10} r$, $\theta$, and $\varphi$ coordinates that spans $0.15\le\log_{10}(r/\rg)\le5, 0\le\theta\le\pi, 0\le\varphi\le2\pi$. We provide the numerical resolution of our simulations in Table~\ref{tab:models}.

\subsection{Resolving the magnetorotational instability (MRI)}

Even though our initial magnetic field is not dynamically important, upon the disk formation, the magnetorotational instability is well resolved \citep[MRI;][]{mri_paper} in all models (except for \pBz). In contrast, when MRI is not sufficiently resolved, it can lead to a runaway density buildup at the inner boundary of the disk, leading to spuriously low values of $\ye$. Our hydrodynamic model (\pBz) is a prime example of this issue on a global scale \citep[see also][for a similar trade-off in hydrodynamic flows]{Dean2024}; locally, this can occur if the unresolved region is large enough.

\subsection{Equation of state}\label{sec:eos}

For the EOS, we adopt Helmholtz EOS, following the approach of \cite{timmes_accuracy_2000}, where thermodynamic quantities such as pressure, specific internal energy, and entropy are computed as the sum of contributions from an ideal gas of ions, radiation, and electrons/positrons of arbitrary degeneracy:
\begin{equation}
	P_{\rm total} = P_{\rm ions} + P_{\rm radiation} + P_{e^\pm}
\end{equation}
The first two components on the right-hand side are calculated analytically. For the degeneracy pressure, we use a precomputed tabulated form of the Helmholtz free energy and its derivatives as functions of density and temperature. The table is spaced logarithmically in density (541 points, $\rho\in[10^{-12},10^{15}]\,\rm{g\,cm}^{-3}$) and temperature (201 points, $T\in[10^{3},10^{13}]\,\rm{K}$). It is linear in the electron fraction ($\ye \in [0,1]$). We note that at densities approaching nuclear saturation density ($\sim 10^{14} \rm{g/cm^3}$), this EOS is no longer valid, since degeneracy pressure is not the main source of pressure in this regime. In practice, in our simulations, densities do not exceed $\approx 10^{13} \rm{g/cm^3}$, well within the validity range of the Helmholtz EOS.

\subsection{Neutrino-transport implementation}\label{sec:nu_transport}

Neutrinos play a crucial role in setting the composition of the accretion disk and its outflows. The degree of neutron richness, expressed through electron fraction $\ye$, will determine whether $r$-process occurs in the outflows, alongside the entropy and the expansion timescale \citep[e.g.,][]{Lippuner2015}. Therefore, we need accurate neutrino transport to model the compositional evolution of the ejected material. In this work, we implement an M1 scheme for the neutrino transport, following \cite{foucart_post-merger_2015,foucart_impact_2016,Sadowski2014} and \cite{mckinney_three-dimensional_2014}, where neutrino-matter interactions are included via a lookup table \citep[Nulib,][]{oconnor_open-source_2015}. Here, we present a brief overview of the scheme and its implementation. We provide a more detailed description of the implementation details, numerical tests, and calculations of the neutrino and antineutrino properties in Appendix \ref{sec:nu_transport_full}. 

We evolve three neutrino species $(\nu_{\rm e}, \bar{\nu}_{\rm e}, \nu_{\rm x})$, where the latter is the combination of four species $(\nu_{\mu}, \bar{\nu}_{\mu}, \nu_{\tau}, \bar{\nu}_{\tau})$. For a single species, we write the neutrino energy-momentum tensor as
\begin{equation}
\begin{split}
    R^{\mu\nu} &= E n^\mu n^\nu + F^\mu n^\nu + F^\nu n^\mu + P^{\mu\nu} \quad \rm (Lab \ frame) \\
    &= J u^\mu u^\nu + H^\mu u^\nu + H^\nu u^\mu + L^{\mu\nu} \quad \rm (Fluid \ frame) \\
    &= \frac{1}{3} E_R \left (4 u_R^\mu u_R^\nu + g^{\mu\nu} \right) \quad \rm (Radiation \ frame)
\end{split}
\end{equation}
where $u^{\mu}$ is the 4-velocity and $g^{\mu\nu}$ is the contravariant metric; and $(E, F^\mu, P^{\mu\nu})$ and $(J, H^\mu, L^{\mu\nu})$ are the moments of the neutrino distribution function (neutrino energy, fluxes, and stress tensor) measured by an observer in the lab (coordinate) and the fluid frames, respectively. We introduce a ``radiation'' frame that moves at 4-velocity, $u^{\mu}_{\rm R}$, whose observer measures the energy density, $E_{\rm R}$. 
The evolution equations come from the energy-momentum conservation equation:
\begin{equation}
\begin{gathered}
    \nabla_{\mu} \left( T^{\mu}_{\nu} + \sum_{s=\nu_e, \bar{\nu}_e, \nu_x} R^{\mu}_{\nu, s} \right) = 0 \\
    T^{\mu}_{\nu} = \left( \rho+u_{\rm g}+P_{\rm g}+b^2 \right) u^{\mu} u_{\nu} + \left( P_{\rm g}+b^2/2 \right) \delta^{\mu}_{\nu} - b^{\mu} b_{\nu}
\end{gathered}
\end{equation}
where $\rho$ is the gas density and $u_{\rm g},P_{\rm g}$ are the gas internal energy and pressure, all measured in the rest frame of the gas. Here, $b^{\mu}$ is the contravariant magnetic field 4-vector (and $b^2=b^{\mu}b_{\mu}$).

To close the evolution equations, we assume a closure relation expressing the stress tensor in terms of the energy and flux densities. In our work, we follow an approach by \cite{Sadowski2014} and \cite{mckinney_three-dimensional_2014}, where we adopt a closure by \cite{Levermore1984}, assuming that there exists a ``radiation'' frame, in which neutrino radiation is isotropic. Then, the lab frame moments can be found by boosting the radiation stress-energy tensor from the radiation frame to the lab (coordinate) frame (with velocity $-u^{\mu}_{\rm R}$).

\section{Simulation results}\label{sec:results}

\begin{figure}[hbtp]
	\centering
	\includegraphics[width=0.95\columnwidth]{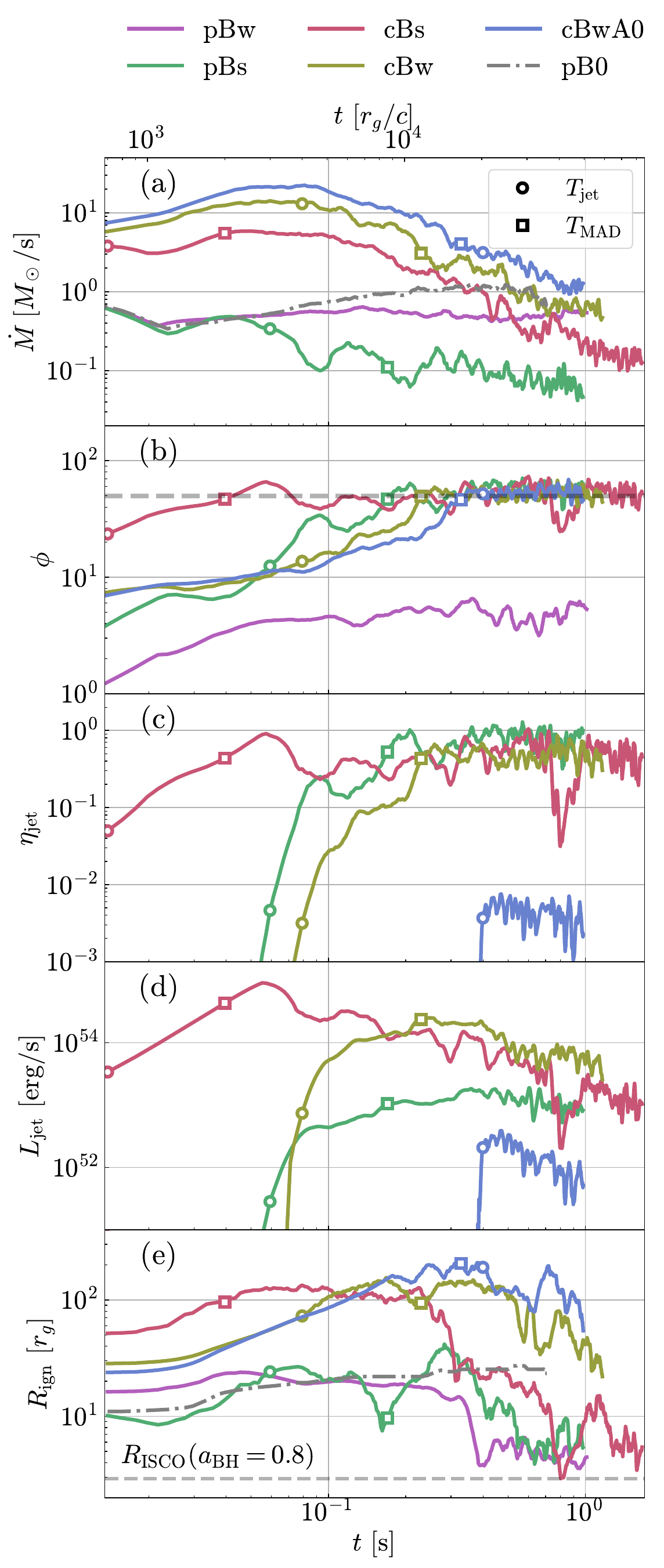}
	\caption{Models with flat density cores (cB) develop large values of mass accretion rate, $ \dot{M} \sim 10\,\msun\,\s^{-1} $ (panel (a)), and dimensionless magnetic flux [panel (b)] that are favorable for the production of neutron-rich material and outflows (see Fig.~\ref{fig:ye_rign_rt}). Once the dimensionless magnetic flux exceeds the critical value, $ \phi \sim 10 $ (panel (b)), the jet-launching efficiency (panel (c)) and power (panel (d)) surge. Spurred by high $\dot M$, the ignition radius, $ R_{\rm ign} $ (panel (e)), initially rises but drops as $ \phi $ increases (panel (b)) and obstructs $ \dot{M} $ (panel (a)). Circles and squares mark $ t_{\rm jet}$  and $t_{\rm MAD} $ of each model, respectively. To reduce clutter, we smoothed all quantities over $\Delta t=10^3 \rg/c=0.02$~s.}
	\label{fig:vs_time}
\end{figure}

Figure~\ref{fig:vs_time}(a) illustrates the time evolution of mass accretion rate measured at $r=5\rg$ (to avoid contamination by density floors at smaller distances) across all models. The mass density profile of the innermost shells dictates the mass accretion rate. We consider flat core density profiles with $ \alpha_{\rm p} = 0 $ in models \cBs, \cBw, and \cBwz, as expected at the initial time of collapse in GRB progenitors \citep[e.g.,][]{Woosley2006}. However, in our simulations, $t=0$ marks the time of BH formation, which is preceded by a few seconds long proto-NS phase \citep[e.g.,][]{Aloy2021}. During this phase, the innermost shells exhibit a free-fall radial density profile, $ \alpha_{\rm p} = 1.5 $, which we adopt in the models \pBs, \pBw, and \pBz\, \citep{halevi_density_2023}. To maintain the same total mass in the star, these models feature higher density values at the center than those with $ \alpha_{\rm p} = 0 $.

Higher central densities lead to higher mass accretion rates that evolve as $ \dot{M} \sim t^{1-2\alpha_{\rm p}/3}$ \citep[see][]{gottlieb_black_2022}. Thus, in models with a flat core (models \cBw, \cBs, and \cBwz), the mass accretion rate is initially very high ($ \dot{M} \sim 10\,\msun\,\s^{-1} $) but begins to decline already at $ t \lesssim 0.1\,\s $. Conversely, models with $ \alpha_{\rm p} = 1.5 $ (models \pBw and \pBz) exhibit a lower but steadier mass accretion rate, which at $ t \gtrsim 1\,\s $ surpasses that of the flat core models. When strong magnetized outflows form (model \pBs; green line), feedback from these outflows hinders accretion, leading to a decline in the mass accretion rate regardless of the density profile. We note that in the model without magnetic fields, \pBz, most of the accretion occurs through the polar funnel and not through the disk, leading to elevated $\dot{M}$ in comparison with models \pBs and \pBw.

Figure~\ref{fig:vs_time}(b), (c) show the dimensionless magnetic flux on the event horizon (Eq.~\ref{eq:phiBH}) and jet-launching efficiency:
\begin{equation}
    \eta_{\rm jet} \equiv \frac{L_{\rm jet}}{\dot{M}}\,,
\end{equation}
where the jet power (or outflow power for model \cBwz) is defined as the power contained in magnetically dominated ($ \sigma = b^2/\rho > 1 $) outflows:
\begin{equation}
    L_{\rm jet} = \int_{r=5\rg} \sqrt{-g} (-T^{r}_{t} - \rho u^{r}) \Big|_{\sigma>1} d\theta d\varphi\,.
\end{equation}
Here, $ T^{r}_{t} $ gives the negative of the radial energy flux density, a component of the mixed stress-energy tensor, $ T^{\mu}_{\nu} $. The last term in parentheses removes the contribution of the rest-mass energy flux from the total energy flux. When the dimensionless magnetic flux exceeds $ \phi \approx 10 $ (panel (b)), a relativistic jet launches, as indicated by a sharp rise in the jet efficiency, $ \eta_{\rm jet} $ (panel (c)). This effect is most pronounced in models \pBs, \cBs, and \cBw, where once $ \phi \gtrsim 10 $ at $ t \lesssim 0.1\,\s $, the jet efficiency surges from $ \eta_{\rm jet} < 10^{-3} $ to $ \eta_{\rm jet} \gtrsim 0.1 $. This result is consistent with an order-of-magnitude estimate of the critical $ \phi $ for jet launching in \S\ref{sec:theory}. Ultimately, the dimensionless magnetic flux reaches the asymptotic MAD level of $ \phi \approx 50 $ (marked by a gray dashed line in Fig.~\ref{fig:vs_time}(b)), where the jet-launching efficiency stabilizes at an order unity. We note that the critical $ \phi $ for magnetized outflow launching depends on the BH spin, as seen in model \cBwz. In this model, the outflows only launch after the MAD state is established, with the launching efficiency remaining low at $ \eta_{\rm jet} \lesssim 10^{-2} $.

\begin{figure}[hbtp]
	\centering
	\includegraphics[width=\columnwidth]{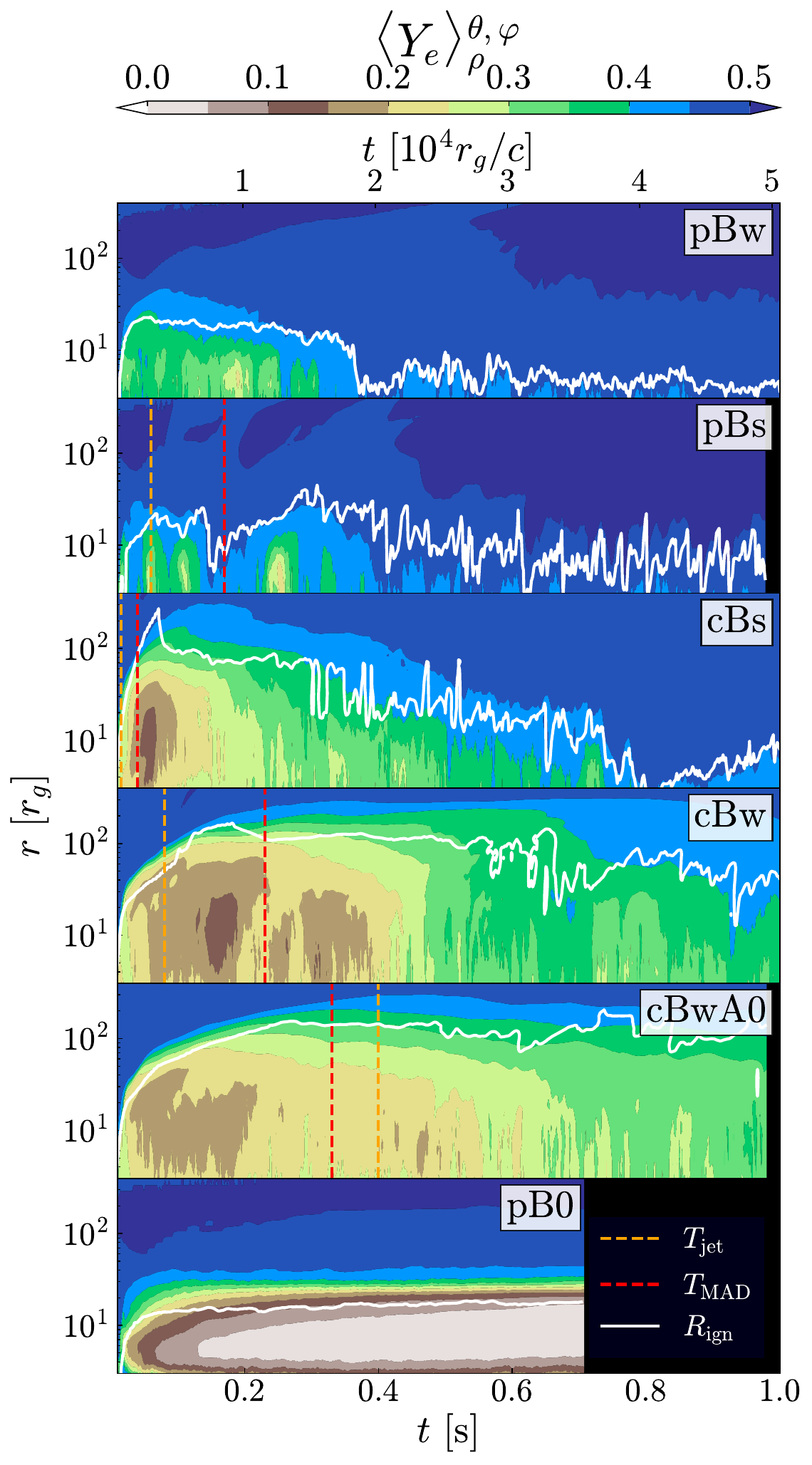}
	\caption{Density-weighted average electron fraction as a function of radius at various times for different models. The regions with neutron-rich material closely track the ignition radii (white lines). The increase in disk magnetic flux leads to jet launching (yellow dashed lines) and a transition to MAD (red dashed lines). This transition signifies the drop in the accretion timescale and increase in the neutrino-cooling timescale due to decreasing densities, which ultimately ends the neutron-rich phase of the disk.}
	\label{fig:ye_rign_rt}
\end{figure}

The product of the accretion power and the jet-launching efficiency determines the jet power, depicted in Fig.~\ref{fig:vs_time}(d). In most models, the combination of high efficiency and high mass accretion rates initially produces extremely powerful jets. However, once $ \phi \approx 50 $, the jet power diminishes over time, following the decline in the mass accretion rate. In model \cBwz, the nonspinning BH system can only launch outflows by extracting the angular momentum from the disk (not the BH). Consequently, the outflow power in this model is lower than that in rest of the magnetized models.

\begin{figure*}[t!]
        \centering
        \includegraphics[width=\textwidth]{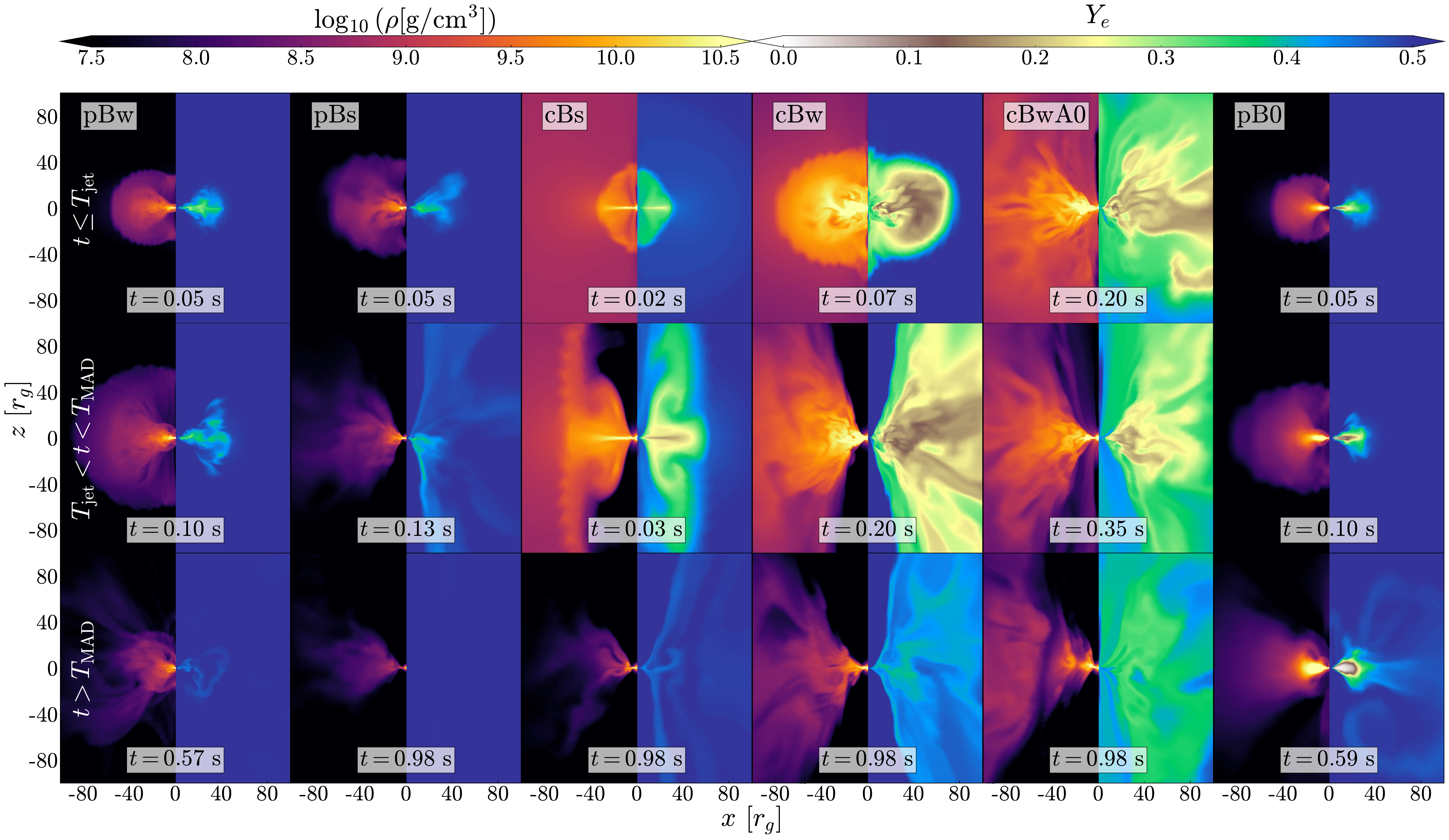}
        \caption{Meridional cuts of the mass density (left-hand side) and electron fraction (right-hand side) for all models (different columns). The top row displays the system before jet launching ($ t < t_{\rm jet} $), featuring neutron-rich disks where $ \ye $ varies with mass accretion rate and disk magnetization. After the launching of jets and disk winds ($ t_{\rm jet} < t < t_{\rm MAD} $), neutron-rich material is ejected from the disk midplane and carried by the shocked cocoon that engulfs the jets. In the bottom row, magnetic fields are dynamically important ($ t > t_{\rm MAD} $), shortening the accretion timescale and reducing the disk densities, leading to a growing neutrino-cooling timescale, which leads to disk deneutronization and the cessation of neutron-rich outflows.}
        \label{fig:megafigure}
\end{figure*}

Fig.~\ref{fig:vs_time}(e) displays the time evolution of the radius within which neutrino cooling is efficient, $ R_{\rm ign} $. This radius increases as the cooling timescale grows and the accretion timescale shortens. In our magnetized simulations, the accretion timescale remains small due to efficient angular momentum transport, driven by large-scale magnetic fields. For models with mass accretion rates below $\dot{M}\sim 1\,\msun\,\s^{-1}$ (\pBs and \pBw), cooling is inefficient due to lower density, resulting in $R_{\rm ign}\sim 10\rg$. Conversely, in models with very high mass accretion rates (\cBw, \cBw, and \cBwz), increased density leads to reduced cooling timescales and a larger $R_{\rm ign}\sim 100\rg $. In all cases, the accretion timescale plummets once the MAD state is established. Combined with the decrease in mass accretion rates, which reflect lower disk densities, which signify longer cooling timescales (see Fig.~\ref{fig:timescales_disk}), $R_{\rm ign}$ inevitably drops. Hence, the MAD state marks the transition from the neutron-rich disk phase to the neutron-poor disk phase, accompanied by an increase in $ \ye $.

Figure~\ref{fig:ye_rign_rt} demonstrates the role of $ R_{\rm ign} $ in estimating the electron fraction in the disk, delineating the density-weighted angle-averaged electron fraction $ \langle \ye \rangle_{\rho}^{\theta,\varphi} (r, t) $  (Eq.~\ref{eq:avg_def}) as a function of radius and time. The definition of $ R_{\rm ign} $ in Eq.~\eqref{eq:Rign} is supported by a good agreement between $R_{\rm ign}$ (solid white line in Fig. \ref{fig:ye_rign_rt}) and the average radial extents of the neutron-rich material in the disk across all models. By the time of jet (outflow) launching (yellow vertical dashed lines in models \pBs, \cBs, \cBw, and \cBwz), some disk material is expelled via disk winds. This ejection contributes to a rise in the average $ \ye $. The onset of the MAD state (red vertical dashed lines) marks the end of the neutron-rich disk state. If the magnetic flux on the BH is insufficient for outflow launching, the neutron-rich material is swiftly accreted (as in model \pBw), or a long-lived neutron-rich disk is established (model \pBz). 

\begin{figure*}[hbtp]
        \centering
        \includegraphics[width=\linewidth]{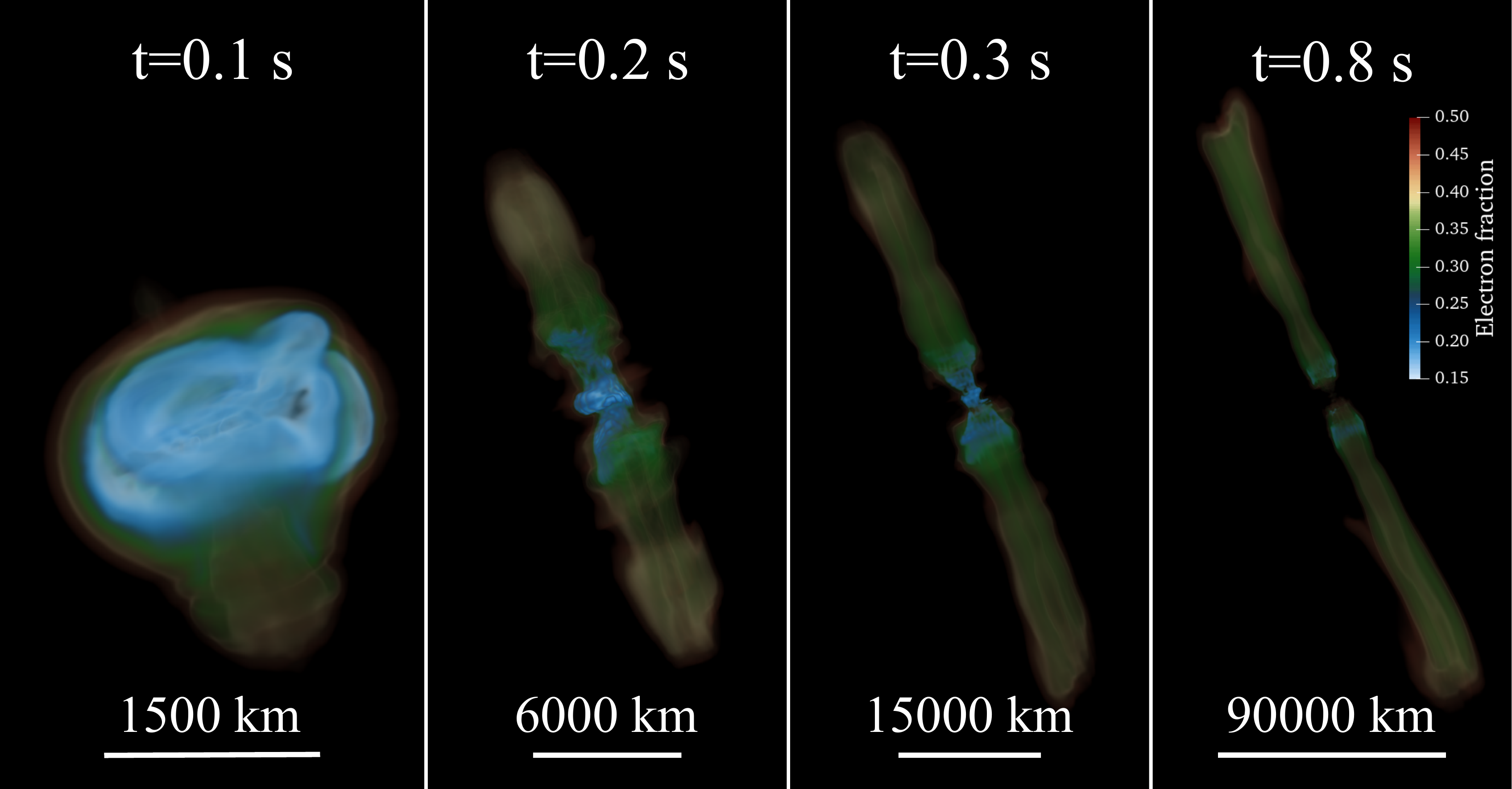}
        \caption{3D rendering of the $ \ye $ in the ejecta of model \cBw, at different times. Around the time of jet launching (left panel), the disk is neutron rich (low $\ye$). At $t=0.2$ s, the jet and neutron-rich winds are launched, and their interaction forms a neutron-rich cocoon. As the cocoon interacts with the neutron-poor gas in the star, significant mixing occurs, increasing the cocoon electron fraction (middle panels). At later times (right panel), the disk density drops significantly, resulting in higher $ \ye $, while the outflows continue to mix with the star material, further raising $ \ye $ in the cocoon. An animation of the evolution of $\ye$ in the ejecta of model \cBw\, is available \href{https://youtu.be/SDhWVtKh338}{here}. The animated 3D rendering is similar to the static version, but it uses a different color table and jet orientation. The animated version of the \cBw\, model proceeds from $t =0.0$ to $1.02$ s over the course of the 54 s movie.}
        \label{fig:3dmap}
\end{figure*}

Figure~\ref{fig:megafigure} shows meridional 2D slices of mass density (left-hand side of each panel) and electron fraction $\ye$ (right-hand side). The first row shows the disk evolution prior to jet launching, at $ t < t_{\rm jet} $. Jet (outflow) launch time $t_{\rm jet}$ is used to mark the moment when outflow launching efficiency exceeds $\eta_{\rm jet} \gtrsim 10^{-3}$ (or when $ \phi \lesssim 10$) and is used for illustrative purposes (as opposed to $t_{\rm MAD}$ defined by the moment when $\phi \approx 50$). In this phase, $ \dot{M} > \dot{M}_{\rm ign} $, which results in the electron fraction dropping below $ \ye = 0.5 $. The specific $ \ye $ value is governed by the ratio of $ t_{\rm cool}/t_{\rm acc} $, which is, in turn, dictated by the mass accretion rate and magnetic flux. In models with very high accretion rates $ \dot{M} \sim 10\,\msun\,\s^{-1} $ (\cBs, \cBw, and \cBwz), where $R_{\rm ign}$ is large, the electron fraction in the disk drops to $ \ye \lesssim 0.2 $. In models with moderate accretion rates and appreciable magnetic fields (\pBw and \pBs), $ R_{\rm ign} $ is smaller and $ \ye \approx 0.3 $. For comparison, in the hydrodynamic model (\pBz; rightmost panels), the absence of MRI-driven turbulent torques results in a longer accretion timescale. Consequently, the gas has enough time to cool efficiently, allowing the electron fraction to drop as low as $ \ye \approx 0.1 $, even with a moderate mass accretion rate. Without powerful outflows and efficient accretion, the disk remains neutron rich until the end of the simulation while slowly spreading as its mass increases, as indicated by the rising $R_{\rm ign}$ (Fig.~\ref{fig:vs_time}(e)). 

The second row of Fig.~\ref{fig:megafigure} depicts the system shortly after jet launching (except for models \pBw and \pBz$\,$ where no jets are present), at $ t_{\rm jet} < t < t_{\rm MAD} $ ($ 10 \lesssim \phi \lesssim 50 $). The growth of $ \phi $ in the disk drives both the jet launching from the BH and emergence of disk winds. The disk winds can eject portions of the low-$ \ye $ material from the disk midplane. As the jets ($\ye\approx 0.5$) shock the neutron-rich winds, they inflate a pressurized neutron-rich cocoon, as shown by the low $ \ye $ values observed in the jet wings. For $ a = 0 $ (model \cBwz), the absence of relativistic jets implies that the magnetized disk winds are the leading force in carrying the neutron-rich ejecta to the stellar edge.

The bottom row of Fig.~\ref{fig:megafigure}, which shows the systems at $ t > t_{\rm MAD} $, depicts the transition from a neutron-rich to a neutron-poor disk once a MAD state is reached. Namely, the increased magnetic flux and reduced mass accretion rate lower $ R_{\rm ign} $, halting the production of neutron-rich material in the disk. Concurrently, the electron fraction in the cocoon rises over time due to two factors. First, the deneutronization of the disk prevents the cocoon from replenishing its neutron-rich gas. Second, the interaction between the cocoon's bulk and the infalling neutron-poor gas increases its $ \ye $. Consequently, while the jet launching marks the efficient ejection of neutron-rich material from the disk, it also signifies the onset of increasing $\ye$. In model \pBw, where the amount of the magnetic flux is insufficient to launch outflows, neutron-rich material remains in the disk and eventually accretes onto the BH.

Figure~\ref{fig:3dmap} summarizes the evolution of the outflow composition with a 3D visualization of model \cBw\, over time. In the initial stage, just before the jet launching (left panel), the disk is characterized by a low electron fraction (blue), indicating a neutron-rich environment. The interaction of the jet and surrounding neutron-rich disk winds generates a shocked cocoon of similar composition (blue, green). The cocoon then interacts with the surrounding neutron-poor gas, leading to substantial mixing and an increase in $ \ye $ to values exceeding 0.4 (yellow, red; middle panels). By the later stages (right panel), the density of the disk drops significantly (becoming invisible due to its reduced size), leading to $ \ye \sim 0.3 $. Simultaneously, the ongoing interaction of the outflows with the remaining star material continues to elevate the electron fraction in the cocoon.

\begin{figure}[]
	\centering
	\includegraphics[width=0.9\columnwidth]{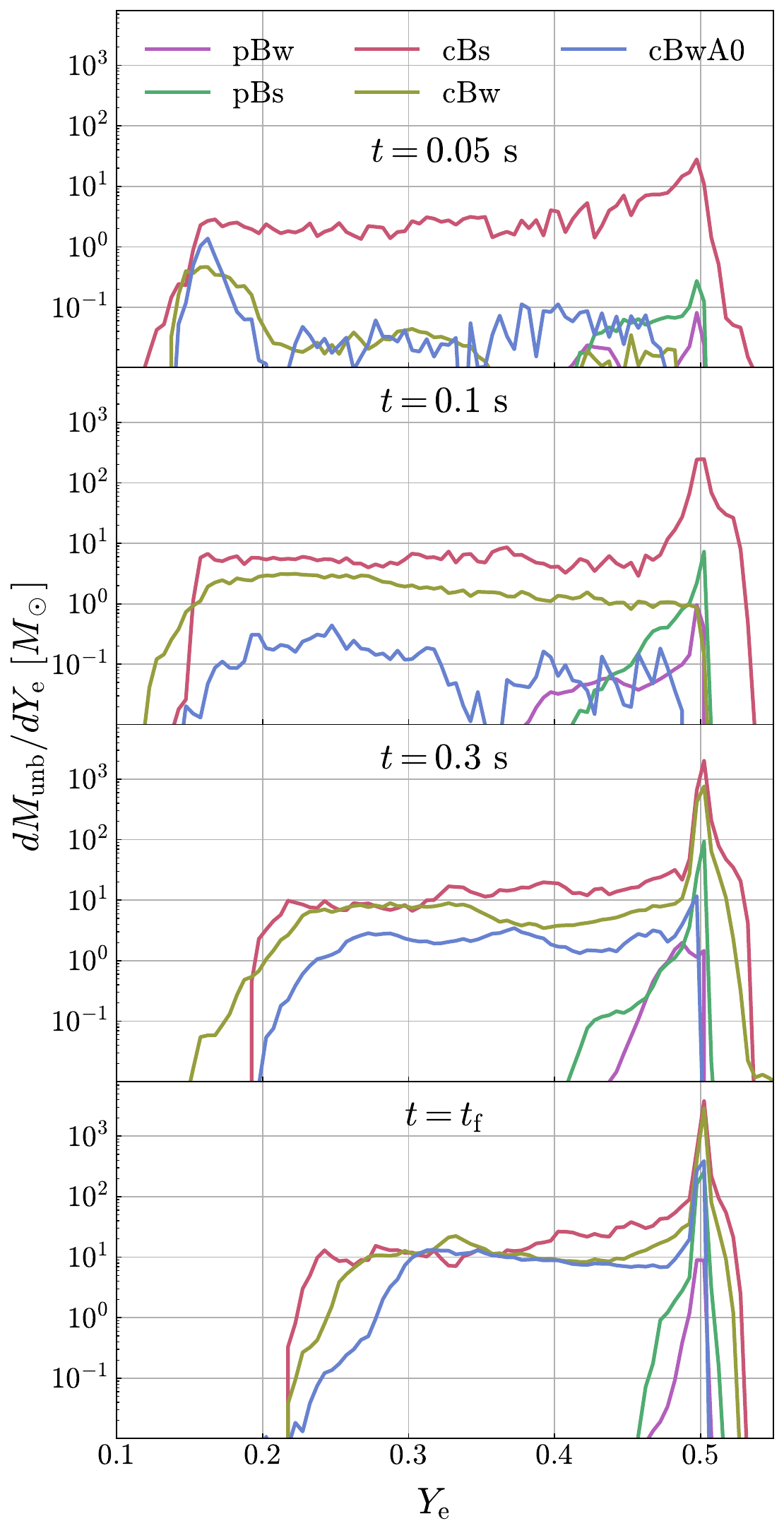}
	\caption{Unbound mass binned by electron fraction in models with significant mass ejection (excluding \pBz) at different times. Unbound mass is characterized according to the Bernoulli criterion in Eq.~\ref{eq:unbound}, where before jet launching, we omit the magnetic field contribution ($b^2/\rho$ term). Stronger fields unbind more low $ \ye $ material, maintaining a nearly flat distribution above the $ \ye $ cutoff. Over time, the mixing of low $ \ye $ outflows with the star raises the lower $ \ye $ cutoff, resulting in an increased mass at higher $ \ye $ values.}
    \label{fig:unb_ye}
\end{figure}

As low-$ \ye $ matter is ejected, weak interactions, particularly $e^{\pm}$ pair captures, freeze-out and the composition is then governed by either neutrino absorption or mixing. Unlike in compact binary mergers, the ejecta continues to interact with the dense star, allowing for ongoing compositional changes through mixing. Figure~\ref{fig:unb_ye} depicts unbound ejecta histograms in $ \ye $ at various times across all models with ejecta. The unbound mass is defined through the Bernoulli criterion:
\begin{equation} \label{eq:unbound}
    -u_{t} \left( 1 + \frac{u_{\rm g} + P_{\rm g} + b^2}{\rho} \right) > 1\,,
\end{equation}
where $ u_t $ is the covariant time component of the 4-velocity and $ u_g $ is the thermal energy density. Before jet launching, $b^2/\rho$ rises above unity along the polar regions where the gas free-falls. However, this magnetized gas ends up being accreted onto the BH. Therefore, we disregard the magnetic field contribution for the unbound criterion at $ t < T_{\rm jet} $.

Models \pBs and \pBw, which have lower mass accretion rates, produce smaller amounts of neutron-rich ejecta. Conversely, models with a flat density core (\cBs, \cBw, and \cBwz) have higher accretion rates and can expel $ \gtrsim 0.1\,\msun $ of roughly uniformly distributed mass in the range of $ 0.15 \lesssim \ye \lesssim 0.5 $ at $ t \approx 0.1\,\s $. Stronger outflows (in model \cBs) expel more neutron-rich ejecta while maintaining a similar flat $ \ye $ distribution within the ejecta. The growth of the magnetic field in the disk halts the production of new low $ \ye $ matter. At the same time, the previously generated neutron-rich ejecta mixes with the $ \ye = 0.5 $ unshocked star, raising the $ \ye $ in the ejecta. The two effects result in a continuous upward shift in the low $ \ye $ distribution cutoff over time, increasing the mass content at the higher $ \ye $ values. For example, at $ t = 0.1\,\s $, the uniform mass distribution of model \cBs$\,$ has a cutoff at $ \ye \approx 0.15 $; by $ t = 0.1\,\s $, the cutoff has shifted to $ \ye = 0.25 $.

While this picture suggests that the ejecta will possess $ \ye \approx 0.5 $ by the time it reaches the stellar surface, there are several caveats about the mixing process. First, the low resolution at large distances from the disk leads to numerical diffusion and mixing that artificially increases the $ \ye $ values. Higher-resolution simulations are anticipated to show a milder evolution of the average $ \ye $ in the ejecta. Second, the neutron rapid capture characteristic timescale of $ t \lesssim 1\,\s $ \citep[e.g.,][]{Lippuner2017} indicates that hydrodynamic mixing over longer timescales will not alter the nuclear composition. The velocity of low $\ye$ outflows can reach a fraction of $c$, and the time dilation effect can effectively elongate the nucleosynthesis timescale. Models \cBs, \cBw, and \cBwz\,exhibit $ M(\ye<0.3) \approx 1\,\msun $ at $ t \approx 0.6\,\s $ after the low $ \ye $ matter was ejected. This implies that the low $ \ye $ ejecta has enough time to synthesize $r$-process nuclei, including lanthanides at $ \ye \lesssim 0.25 $ \citep[e.g.,][]{Lippuner2015}, which will blend into the supernova ejecta, increasing the effective opacity and reddening the emission signature (e.g., \citealt{Barnes&Metzger22,Barnes&Duffell23}).

\section{Summary}\label{sec:summary}

In this Letter, we present the first global GRMHD simulations of collapsar explosions incorporating M1 neutrino transport. We demonstrate that strongly magnetized disks eject massive ($ M_{\rm ej} \approx 1\,\msun $), neutron-rich ($ \ye < 0.25 $) outflows at extreme mass accretion rates ($ \dot{M} \sim 10\,\msun\,\s^{-1} $). We also show that strong magnetic fields, necessary for outflow launching in collapsars, shorten the accretion timescales in the disk, simultaneously growing the neutrino-cooling timescales due to decreasing densities and thus suppress further disk neutronization.

As the system approaches a MAD state, disk outflows unbind neutron-rich material from the disk midplane, and a Kerr BH also launches twin relativistic jets. The interaction between the jets and the disk outflows forms a neutron-rich cocoon that mixes with the neutron-poor gas in the star. However, the field growth in the disk reduces the accretion timescale below that of weak interactions, rendering disk neutronization and jet launching mutually exclusive. Despite this, the outflows still transport some of the neutron-rich material synthesized in the disk, suggesting a potential mechanism for powering $r$-process ejecta from collapsars. A similar trade-off between unbinding neutron-rich material from the disk and disk neutronization was observed in hydrodynamic collapsar simulations \citep{Just2022,Dean2024}. In these simulations, the absence of a magnetized outflow delays the rapid expansion of the shock until the disk transitions to the advection-dominated accretion flow phase due to the decreasing disk density.

In our models, neutron-rich disk formation and ejection occur in a single episode. However, in nature, multiple cycles might take place if the magnetic flux in the envelope is more stochastic, or if the angular momentum profile allows for the intermittent formation and destruction of the disk. Such repeated episodes could significantly increase the total amount of neutron-rich ejecta. Furthermore, if an intermittent jet progressively unbinds a larger fraction of the star, the neutron-rich ejecta could propagate outward with minimal mixing, similar to compact binary mergers, thereby enhancing the efficiency of heavy $r$-process nucleosynthesis. Alternatively, if the BH inherits its magnetic field from the proto-NS \citep{Gottlieb2024}, it could enable jet launching while maintaining moderate disk magnetization, potentially allowing for simultaneous disk neutronization and neutron-rich ejecta production. We plan to explore these scenarios in future work.

In our simulations, we initiate exceptionally strong magnetic fields to resolve the MRI within the disk. This approach leads to high disk inflow velocities, forcing our models to maintain extremely high mass accretion rates of $ \dot{M} \sim 10\,\msun\,\s^{-1} $ to facilitate disk neutronization. These high mass accretion rates also drive extreme jet energies of $ 10^{53}\,{\rm erg} \lesssim E_{\rm jet} \lesssim 10^{54}\,{\rm erg} $. While very atypical, the lower-energy end of such extreme GRBs exists in nature \citep[][]{Burns2023,boat2023}, hinting at a possible connection of extremely bright GRBs with large neutron-rich ejecta mass. The high mass accretion rates will also cause the BH to double its mass within the collapse time, leading to a substantial spin-down \citep{beverly_spindown,jon_spindown}, which complicates our assumption of static spacetime. Similarly, although our simulations produce up to $\mej \approx\msun$ of ejecta with $\ye<0.25$ within the freeze-out time, these results are influenced by the magnetic field strength and density profile of the collapsing star, which yield exceptionally high accretion rates and jet power in our models. Moreover, the final $ \ye $ may change depending on what neutrino-matter interactions are taken into account in the neutrino transport; therefore, the outcomes could be different if more sophisticated neutrino-transport calculations are performed.

In summary, our findings imply that only certain special collapsar progenitors may contribute to the Galactic $r$-process enrichment. Hence, reduced estimates of $r$-process production from collapsars could better describe the observed Galactic abundances \citep[see e.g.,][for estimates]{Rastinejad2024}. The precise mass and composition of neutron-rich ejecta across different progenitors remain uncertain. To address this, future studies must prioritize higher-resolution simulations that permit weaker magnetic fields, more moderate accretion rates, and jet powers, allowing for the formation of neutron-rich ejecta under more realistic conditions. Stellar evolution models can provide such initial conditions for our models, which we aim to explore in the follow-up works. Furthermore, our current simulations lack crucial nucleosynthesis calculations with passive tracers to capture the thermodynamic evolution of the ejecta and do not include postprocessing via nuclear reaction networks. These features are essential for determining whether the neutron-rich ejecta can undergo $r$-process on seed nuclei before mixing with high-electron-fraction infalling material.

\begin{acknowledgments}
\section*{Acknowledgements}
We thank Irene Tamborra, Rodrigo Fern\'andez, Philipp M{\"o}sta, and Daniel Kasen for useful comments and discussions. We thank the anonymous reviewer for their feedback that led to improvements throughout the manuscript. D.I. thanks Aretaios Lalakos and Nick Kaaz for fruitful discussions.
D.I. is supported by Future Investigators in NASA Earth and Space Science and Technology (FINESST) award No. 80NSSC21K1851. %Finesst
O.G. acknowledges the Flatiron Research and CIERA Fellowships. B.D.M. acknowledges support from the National Science Foundation (grant No. AST-2002577) and Simons Investigator grant 727700. The Center for Computational Astrophysics at the Flatiron Institute is supported by the Simons Foundation.
J.J. acknowledges support by the NSF AST-2009884, NASA 80NSSC21K1746 and NASA XMM-Newton  80NSSC22K0799 grants.
This work was performed in part at the Kavli Institute for Theoretical Physics (KITP) supported by grant NSF PHY-2309135.
This work was performed in part at Aspen Center for Physics, which is supported by National Science Foundation grant PHY-2210452.
This research used resources of the National Energy Research Scientific Computing Center, a DOE Office of Science User Facility supported by the Office of Science of the U.S. Department of Energy under Contract No. DE-AC02-05CH11231 using NERSC allocations m4603 (award NP-ERCAP0029085) and m2401. The computations in this work were, in part, run at facilities supported by the Scientific Computing Core at the Flatiron Institute, a division of the Simons Foundation. An award of computer time was provided by the ASCR Leadership Computing Challenge (ALCC), Innovative and Novel Computational Impact on Theory and Experiment (INCITE), and OLCF Director’s Discretionary Allocation programs under award PHY129. A.T. acknowledges support by NASA 
80NSSC22K0031, %fermi GRB grant
80NSSC22K0799, %XMM-Newton neutron stars
80NSSC18K0565 %neutrino francois
and 80NSSC21K1746 %atp pulsars
grants, and by the NSF grants 
AST-2009884, %neutrino factories
AST-2107839, %crossing the chasm, new short GRB grant
AST-1815304, %old short GRB grant
AST-1911080, %accretion grant
AST-2206471, %tidal disruptions
AST-2407475, %collapsar
OAC-2031997. %Frontera travel grant

\end{acknowledgments}

\bibliography{refs}

\appendix
\section{Neutrino-transport implementation details}\label{sec:nu_transport_full}

\subsection{Equations}

We added the M1 neutrino-transport scheme to the \hammer\, code \citep{hamr_paper} that integrates GRMHD equations of motion. In the presence of neutrino radiation, the evolution equations are
\begin{equation} \label{eq:evol_1}
    \nabla_{\mu} \left( T^{\mu}_{\nu} + \sum_{s=\nu_e, \bar{\nu}_e, \nu_x} R^{\mu}_{\nu, s} \right) = 0
\end{equation}
where $T^{\mu}_{\nu}$ is a magnetohydrodynamical stress-energy tensor:
\begin{equation}
    T^{\mu}_{\nu} = \left( \rho+u+p+b^2 \right) u^{\mu} u_{\nu} + \left( p+b^2/2 \right) \delta^{\mu}_{\nu} - b^{\mu} b_{\nu}
\end{equation}
and $R^{\mu}_{\nu,s}$ is a stress-energy tensor of a single species of neutrinos. Here, we evolve three neutrino species $(\nu_{\rm e}, \bar{\nu}_{\rm e}, \nu_{\rm x})$, where the latter is the combination of four species $(\nu_{\mu}, \bar{\nu}_{\mu}, \nu_{\tau}, \bar{\nu}_{\tau})$. We make this choice because the temperatures and neutrino energies in our collapsar simulations are too low and the formation of heavy lepton neutrinos is suppressed \citep[e.g.,][]{foucart_post-merger_2015}. From here on, we imply the summation over all species and drop the index $s$.

In M1 method, we only consider the first two moments of the neutrino distribution function in the so-called ``gray'' approximation, where we consider energy-integrated moments \citep[see][]{Shibata2011}. Generally, the neutrino stress-energy tensor can be expressed through moments in the lab (coordinate) or fluid (comoving) rest frames:
\begin{equation}
\begin{split}
    R^{\mu\nu} &= E n^\mu n^\nu + F^\mu n^\nu + F^\nu n^\mu + P^{\mu\nu} \quad \rm (Lab \ frame) \\
    &= J u^\mu u^\nu + H^\mu u^\nu + H^\nu u^\mu + L^{\mu\nu} \quad \rm (Fluid \ frame)
\end{split}
\label{eq:Rmunu}
\end{equation}
where $E,\, F^{\mu},\, and P^{\mu\nu}$ are the energy, flux, and stress tensor of the neutrino radiation measured by an observer in an inertial (zero angular momentum observer, or ZAMO) frame and $n^{\mu}$ is the 4-velocity of the ZAMO ($n_{\nu}=\left( -g^{tt} \right)^{-1/2} \delta^{t}_{\nu}$); similarly $J,\, H^{\mu},\, and L^{\mu\nu}$ are the energy, flux, and stress tensor of the neutrino radiation measured by an observer in the fluid rest frame, and $u^{\mu}$ is the fluid 4-velocity. Only the first two moments (energy and flux) are evolved, and to close the set of equations, we need to express the stress tensor (Eq.~\eqref{eq:Rmunu}) in terms of the first two moments (closure relation). 

\subsection{Closure relation}

In this work, we adopt a closure following \cite{Levermore1984}, where the main assumption is that there exists a ``radiation'' frame, an orthonormal frame in which the neutrino radiation is isotropic and satisfies the Eddington closure, which in the fluid frame gives
\begin{equation}
    L^{\mu\nu} = \frac{1}{3} J (g^{\mu\nu} + u^{\mu} u^{\nu})
\end{equation}
Following \cite{Sadowski2014} and \cite{mckinney_three-dimensional_2014}, we consider the decomposition of $R^{\mu}_{\nu}$ in terms of moments in the so-called ``radiation" frame:
\begin{equation}
    R^{\mu}_{\nu} = \frac{4}{3} E_{\rm R} u_{\rm R}^{\mu} {u_{\rm R}^{\phantom{\mu}}}_{\nu} + \frac{1}{3} E_{\rm R} \delta^{\mu}_{\nu}
\end{equation}
where $E_{\rm R}$ is the energy of the neutrino radiation measured by an observer in the ``radiation'' rest frame, and $u^{\mu}_{\rm R}$ is the 4-velocity of the ``radiation'' frame.

This closure reproduces the optically thin and thick limits relatively well. However, unsurprisingly, it breaks down in the presence of colliding beams because of the key assumption that a single radiation frame exists. Since only one direction -- that of the boost -- is distinguished, in case of intersecting beams, the scheme approximates the radiation having the mean direction \citep{Sadowski2014}. Alternative analytic closures suffer from similar problems \citep[see][for an overview]{Murchikova_closures}.

\subsection{Neutrino evolution equations}

We rewrite the evolution equation \ref{eq:evol_1} as
\begin{equation} \label{eq:evol_2}
    \nabla_{\mu} T^{\mu}_{\nu} = - \nabla_{\mu} R^{\mu}_{\nu} = G_{\nu}
\end{equation}where $G_{\nu}$ is the source term (or 4-force) that encapsulates the neutrino-matter interactions:
\begin{equation}
    G_{\nu} = -\eta u_{\nu} + \kappa_{\rm a} J u_{\mu} + \left( \kappa_{\rm a} + \kappa_{\rm s} \right) H_{\mu}
\end{equation}
where $\eta, \kappa_{\rm a}, and \kappa_{\rm s}$ are gray (energy-integrated) emissivities, absorption, and scattering opacities, described in \ref{sec:opacities}. Expressed in conservative form, Eq.~\ref{eq:evol_2} becomes
\begin{equation}
\begin{split}
    \partial_t \left( \sqrt{-g} \rho u^t \right) + \partial_i \left( \sqrt{-g} \rho u^i \right) &= 0, \\
    \partial_t \left( \sqrt{-g} T^t_{\nu} \right) + \partial_i \left( \sqrt{-g} T^i_{\nu} \right) &= \sqrt{-g} \Gamma^{\lambda}_{\nu \kappa} T^{\kappa}_{\lambda} + \sqrt{-g} G_{\nu} \\
    \partial_t \left( \sqrt{-g} R^t_{\nu} \right) + \partial_i \left( \sqrt{-g} R^i_{\nu} \right) &= \sqrt{-g} \Gamma^{\lambda}_{\nu \kappa} R^{\kappa}_{\lambda} - \sqrt{-g} G_{\nu}
\end{split}
\end{equation}
In addition, we also evolve the neutrino number density to accurately track the composition evolution following the approach proposed by \cite{foucart_impact_2016} and \cite{radice_new_2022}. The neutrino number density evolution equation is
\begin{equation}
    \nabla_{\mu} N^{\mu} = \sqrt{-g} \left( \eta_{\rm N} - \kappa_{\rm N} \hat{N} \right)
\end{equation}
where $\hat{N}$ is the neutrino number density in the fluid frame $\hat{N} = - N^{\mu} u_{\mu}$, $\eta_{\rm N}, \kappa_{\rm N}$ are the neutrino number emissivity and absorption coefficients (see \ref{sec:opacities}). We adopted the form
\begin{equation}
    N^\mu = \hat{N} \left( u^\mu + \frac{H^\mu}{J} \right)
\end{equation}
for the neutrino number current under the assumption that the neutrino number and energy flux are aligned, which is exact if neutrinos had a single energy \citep{radice_new_2022}. The composition evolution is then (from lepton number conservation)
\begin{equation}
    \partial_t \left( \sqrt{-g} \rho u^t \ye \right) + \partial_i \left( \sqrt{-g} \rho u^i \ye \right) = -\text{sign}\left( \nu_s \right) \sqrt{-g} \left( \eta_{\rm N} - \kappa_{\rm N} \hat{N} \right)
\end{equation}
where $\rm{sign}\left( \nu_s \right)$ is $1$ for $\nu_{\rm e}$, $-1$ for $\bar{\nu}_{\rm e}$ and $0$ otherwise.

\begin{figure}[hbtp]
	\centering
	\includegraphics[height=1.5in]{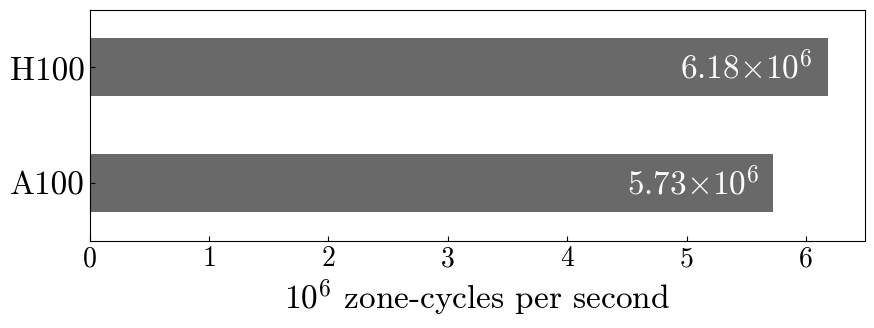}
	\caption{The benchmark of \textsc{$\nu$h-amr}, performed using the simulation initialized with the setup of the model \cBw\, on the grid with total size $200\times 200\times 200$ on a single GPU ($2\times 2\times 2$ blocks with $100\times 100\times 100$ cells per block.)}
	\label{fig:perf_benchmark}
\end{figure}

\subsection{Numerical implementation}

In our simulations the neutrino-matter interactions are weak (radiative efficiency is $\sim$ few percent); thus it is reasonable to separate the evolution of the MHD fluid and neutrino radiation. The evolution algorithm is the following:
\begin{enumerate}
    \item Evolve $T^{\mu}_{\nu}$ and $R^{\mu}_{\nu}$ over the time step $\Delta t$ (subject to Courant condition) by integrating the fluxes and gravitational source terms explicitly without taking into account neutrino-matter interactions (the source term). For details of computation of fluxes in optically thin and thick regions, see \ref{sec:flux_nu}.
    \item Compute the neutrino-matter interaction step, where the source term is treated implicitly. Once the neutrino quantities are updated, compute the neutrino number evolution equation implicitly, using the updated quantities (see \ref{sec:implicit_step}).
    \item Using the updated conserved quantities, convert them to primitive quantities. For MHD quantities, we used a two-dimensional Newton-Raphson root-finding method \citep{Noble2006} and backup entropy-based inversion methods \citep[see][for more details]{hamr_paper}. Neutrino quantities are inverted via the method described in \ref{sec:con2prim}.
    \item Figure~\ref{fig:perf_benchmark} shows the performance of the code on A100 and H100 NVIDIA GPUs. The benchmark is performed using the simulation initialized with the setup of the model \cBw on the grid with total size $200\times200\times200$ on a single GPU ($2\times2\times2$ blocks with $100\times100\times100$ cells per block.).
\end{enumerate}

\subsubsection{Fluxes and characteristic speeds} \label{sec:flux_nu}

We use the Lax-Friedrichs flux for neutrino radiation, for which we need to compute the maximal characteristic wavespeed $a$:
\begin{equation}
    \textbf{F} = \frac{\textbf{F}_R + \textbf{F}_L}{2} - \frac{a (\textbf{U}_R - \textbf{U}_L)}{2}
\end{equation}
where $\textbf{F}_{L,R}, and \textbf{U}_{L,R}$ are fluxes and conserved variables computed at the left and right cell interfaces, respectively, from the reconstructed values of primitive variables using the PPM limiter. We consider both optically thin and thick limits to determine $a$. In the optically thin limit, we take the speed of the radiation characteristic to be $c_{\rm rad}=\pm 1/\sqrt{3}$. In the thick limit, we follow the prescription used in \cite{Sadowski2014} and \cite{mckinney_three-dimensional_2014}, where we take
\begin{equation}
    a_R^i = \min \left( a_R^i, \frac{4}{3\tau^i} \right), \quad a_L^i = \min \left( a_L^i, -\frac{4}{3\tau^i} \right) 
\end{equation}
(here, $a_{R,L}^i$ are right/left going wavespeeds in direction $i$, $\tau^i=\sqrt{g_{ii}}\Delta x^i(\kappa_{\rm a}+\kappa_{\rm s})$ is the optical depth in the cell in direction $i$) to limit the effect of large numerical diffusion.

\subsubsection{Conserved-to-primitive variables inversion} \label{sec:con2prim}

The conserved neutrino radiation quantities $U_{\nu}^{\rm(RAD)} = \sqrt{-g} R^{t}_{\nu}$ have to be converted to primitive quantities $\vec{P}^{\rm(RAD)} = [E_{\rm R}, \Tilde{u}_{\rm R}^{i}]$, where $\Tilde{u}_{\rm R}^{i} = u^{i}_{\rm R} + \left( u^{\mu} n_{\mu} \right) n^i = u^{i}_{\rm R} - \gamma_{\rm R} n^i$, at least once per time step. 
\begin{equation}
    \begin{split}
        U_{\nu}^{\rm(RAD)} &= \sqrt{-g} R^{t}_{\nu} = \sqrt{-g} n^{t} \left( En_{\nu} + F_{\nu} \right) \\
        E &= n_{\mu} n^{\nu} R^{\mu}_{\nu} = -\frac{1}{\sqrt{-g} n^{t}} n^{\nu} U^{\rm(RAD)}_{\nu} \\
        F_{i} &= - n_{\mu} R^{\mu}_{i} = \frac{1}{\sqrt{-g}n^t} U^{\rm(RAD)}_{i}
    \end{split}
\end{equation}
Using the fact that $y \equiv F^2 / E^2$ is only allowed to range between $0$ and $1$, we compute $\gamma_{\rm R}$:
\begin{equation}
    \gamma_{\rm R}^2 = \frac{2-y + \sqrt{4-3y}}{4(1-y)}
\end{equation}
and primitive quantities are recovered via
\begin{equation}
    \begin{split}   
    E_{\rm R} &= \frac{3E}{4\gamma_{\rm R}^2 - 1} \\
    \Tilde{u}_{\rm R}^i &= \left( \frac{4\gamma_{\rm R}^2 - 1}{4\gamma_{\rm R}} \right) \frac{F^i}{E}
    \end{split}
\end{equation}
There are two cases when we have to resort to limiting the radiation primitive variables \citep[see][for 'BASIC' limiter]{mckinney_three-dimensional_2014}. First, in case $E<0$, we reset $E_{\rm R}$ to a floor value and set $\Tilde{u}_{\rm R}^i = 0$. Second, we do not allow $\gamma_{\rm R}>\gamma_{\rm R,\max}$, and if it happens, we rescale $\Tilde{u}_{\rm R}^i$ corresponding to $\gamma_{\rm R,\max}$. We use the value of $\gamma_{\text{R},\max}=50$ in the collapsar simulations.

\subsubsection{Implicit step} \label{sec:implicit_step}

After the fluxes and geometric source terms are applied explicitly, we apply the source term implicitly, where we express $G_{\nu}$ in terms of final conserved quantities:
\begin{equation}
\begin{split}
    U_{\nu}^{\rm(RAD)} &= U_{\nu}^{\rm(RAD)} \Big|_{expl.} - \sqrt{-g} G_{\nu} \Delta t \\
    &= \sqrt{-g} \Delta t \eta u_\mu \\ 
    &- \Big[ (\kappa_{\rm T} n_\mu - 2 \kappa_{\rm s} w u_\mu) u^\nu - \Big(\kappa_{\rm s} u_\mu (p^{ij}u_i u_j - w^2) + \kappa_{\rm T} (w n_\mu + p^j_\mu u_j)\Big) n^\nu - \kappa_{\rm T} w \delta^\nu_\mu \Big] n_t \Delta t U_\nu^{\rm(RAD)}
\end{split}
\end{equation}
Here $\kappa_{\rm T} = \kappa_{\rm a} + \kappa_{\rm s}$ and $w=-n_{\mu} u^{\mu}$, and operating under the assumption that $p^{ij} = P^{ij}/E$ stays constant during this implicit time step. We end up with a $4\times 4$ matrix inversion per species per time step. Finally, number density is evolved using the updated values of the primitive quantities.

\subsubsection{Neutrino-Matter interactions} \label{sec:opacities}

We use the emission, absorption, and scattering coefficients computed via NuLib \citep{oconnor_open-source_2015}. The table includes $\nu_{\rm e}$ absorption on neutrons, $\bar{\nu}_{\rm e}$ absorption on protons, as well as production of $\nu_{\mu}\bar{\nu}_{\mu}$ and $\nu_{\tau}\bar{\nu}_{\tau}$ pairs from $e^{+} e^{-}$ annihilation and nucleon-nucleon Bremsstrahlung. For electron-type neutrinos, we do not include the pair production channels. All inverse reactions are also included, in such a way that the emissivity $\eta$ and absorption opacity $\kappa_{\rm a}$ satisfy Kirchhoff’s law $\eta/\kappa_{\rm a} = B_{\nu}$, with $B_{\nu}$ the blackbody distribution function of neutrinos in equilibrium with the fluid, integrated over the relevant energy bin. We also include in the table elastic scattering on protons and neutrons. Emissivity $\eta$, absorption $\kappa_{\rm a}$ and scattering $\kappa_{\rm s}$ opacities from NuLib are functions of density, temperature, electron fraction, and neutrino energy bin. Then, the gray emissivity is $\bar{\eta}=\sum_{b} \eta(E_b) \Delta E_b$ where $E_b$ and $\Delta E_b$ are the energy bin and bin width, summing over all bins. To obtain the gray absorption opacity, we assume
Kirchhoff's law $\kappa_{\rm a}(E_\nu) B_{\nu}(E_\nu) = \eta(E_\nu)$, such that the energy averaged opacity is \citep{foucart_post-merger_2015}:
\begin{equation}
    \bar{\kappa}_{\rm a}^{\rm (eq)} = \frac{\int_0^{\infty} \kappa_{\rm a}(E_{\nu}) B_{\nu}(E_\nu) dE_\nu}{\int_0^{\infty} B_{\nu}(E_\nu) dE_\nu} \approx \frac{\bar{\eta}}{\sum_b \frac{\eta(E_b)}{\kappa_{\rm a}(E_b)} \Delta E_b}
\end{equation}

The scattering opacity is computed using the same expression. In optically thick regions, this prescription is accurate. In the optically thin regions, we use the fact that the cross sections of the processes used here scale as the square of the average neutrino energy:
\begin{equation}
    \bar{\kappa}_{\rm a,s} = \bar{\kappa}_{\rm a,s}^{\rm (eq)} \frac{T_{\nu}^2}{T_{\rm g}^2}
\end{equation}

The neutrino temperature $T_{\nu}$ is computed approximately using the neutrino number density:
\begin{equation}
    T_{\nu} = \frac{F_2 (\mu_\nu/k_B T_{\rm g})}{F_3 (\mu_\nu/k_B T_{\rm g})} \frac{J}{\hat{N}}
\end{equation}
where $F_k(\zeta)=\int_0^{\infty} dx\, x_k / (1+e^{x-\zeta})$ is the Fermi integral and $\mu_\nu$ is the neutrino chemical potential. 

\section{Numerical tests of the neutrino scheme}

\subsection{Shadow tests}

\begin{figure}[htbp]
    \includegraphics[height=1.3in]{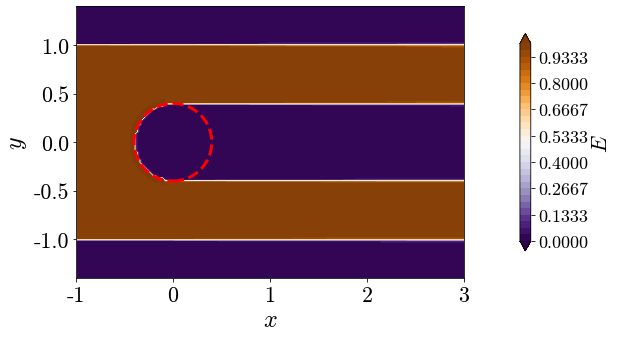}
    \label{fig:opaque_test_beam}
    \includegraphics[height=1.3in]{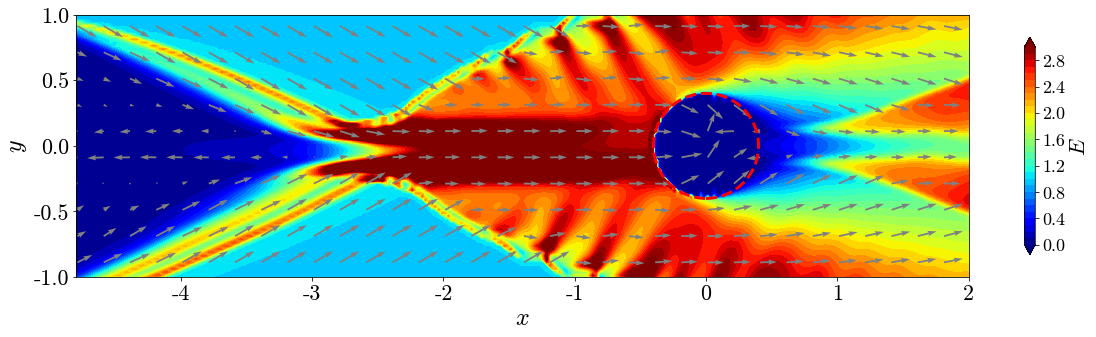}
    \label{fig:opaque_test_twobeams}
    \caption{Lab frame energy density in the radiative shadow tests, where either a single beam (left) or intersecting plane parallel beams (right) shine on an optically thick sphere (red dashed circle), producing a shadow.}
\end{figure}

These tests demonstrate the ability of our scheme to resolve shadows. In flat spacetime, we place an opaque sphere at $(x, y) = (0, 0)$ of radius $R=0.4$ where we set the absorption opacity to $\kappa_{\rm A} = 300$, surrounded by an optically thin medium. 

First, we set a single beam of neutrinos at the left boundary, $x=-1, -1 \leq y \leq 1$, and solve the problem in 2D (Fig.~\ref{fig:opaque_test_beam}). The grid has a range $x \in (-1, 3), y \in (-1.5, 1.5)$ with a resolution of $160\times 120$. Initial radiative energy density is $n_{\mu} n_{\nu} R^{\mu}_{\nu} = E = 1$ and flux density is $F_{x}=0.999998 E$ (corresponding to $\gamma_R=2.5\times 10^5$). Once we evolve the system, we see a sharp shadow form behind the sphere. This setup is particularly favorable for the M1 scheme due to its simple geometry and single illumination source. 

\begin{figure}[hbtp]
	\centering
	\includegraphics[height=2in]{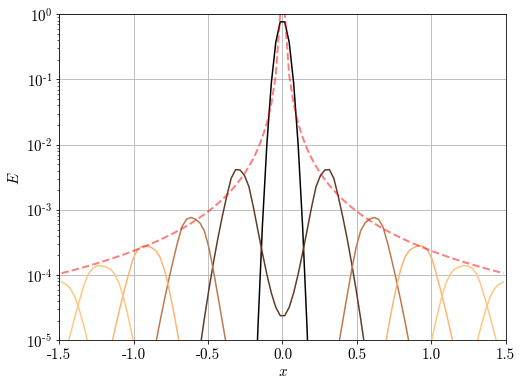}
	\caption{Gaussian pulse propagating in an optically thin medium at different times. The pulse propagates at the speed of light, and its peak follows a $\propto 1/r^2$ profile (dashed red line).}
	\label{fig:pulse_thin_test}
\end{figure}

Next, we set up a test where upper and lower boundaries at $y=-1.5,1.5$ shine planar beams at an inclination of $\theta=\pi/6$ (Fig.~\ref{fig:opaque_test_twobeams}). The grid has a range $x \in (-6, 4), y \in (-1.5, 1.5)$ with a resolution of $400\times 120$. Where the beams overlap, we observe the flux directed along the superposition of the directions of two intersecting beams, and energy density around twice to thrice the initial energy densities of the beams, with higher amplification along the midline of intersection. 

\subsection{Radiative pulse}

In this test, we check how our scheme handles the evolution of a pulse of radiation in the optically thin medium. We set up a Gaussian distribution of the radiative energy density $E = \exp{\left( -r^2/w^2 \right)}$, with $w = 0.05$ (Fig.~\ref{fig:pulse_thin_test}). The grid is a 3D cube with $x, y, z \in (-1.5, 1.5)$ with a resolution of $150^3$. The pulse propagates radially with the speed of light, and its peak magnitude reduces as $\propto 1/r^2$ (red dashed line), as expected in 3D.

\subsection{Radiative sphere}

\begin{figure}[htbp]
    \centering
    	\includegraphics[height=2in]{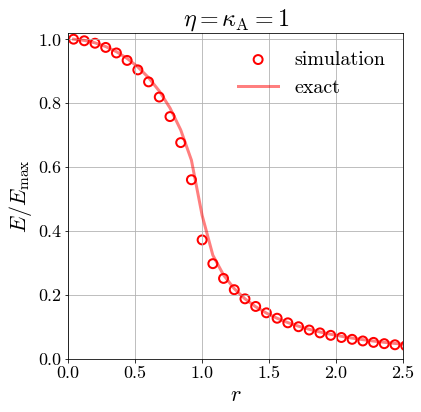}
    	\includegraphics[height=2in]{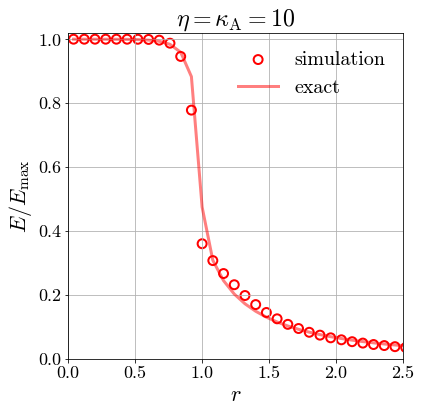}
    \caption{The distribution of energy density in a radiative sphere with radius $r=1$, for two values of absorption opacities. The emission and absorption are in equilibrium inside the sphere, and the medium outside the sphere is optically thin. The steady-state profiles (red circles) show a good agreement with the analytical solution (red lines).}
    \label{fig:rad_sphere_test}
\end{figure}

\begin{figure}[htbp]
        \centering
    	\includegraphics[height=2.5in]{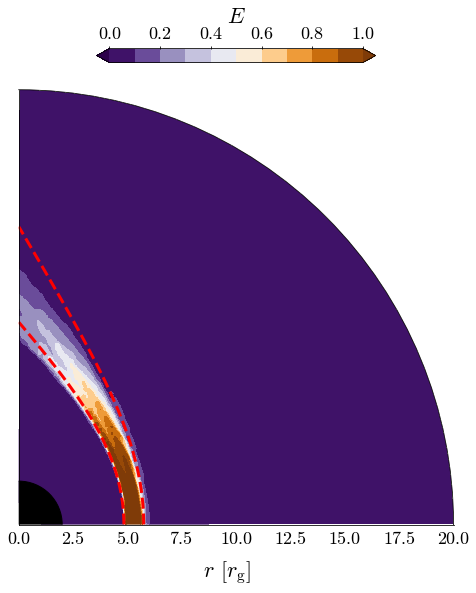}
    	\includegraphics[height=2.5in]{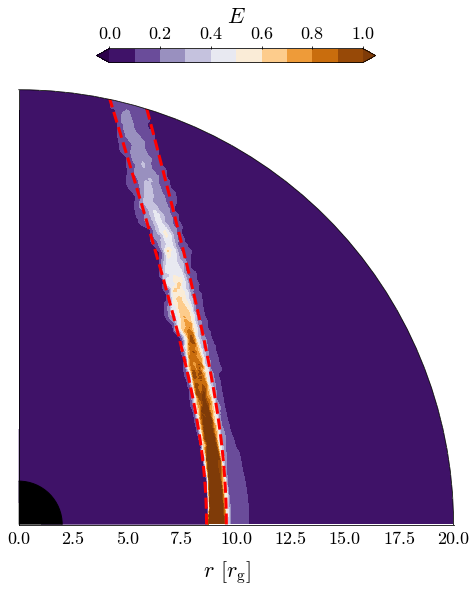}
    	\includegraphics[height=2.5in]{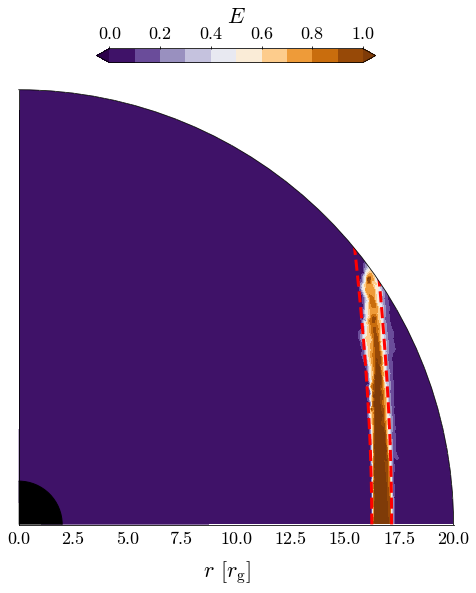}
    \caption{Lab frame energy density of the beams propagating in the curved spacetime of a Schwarzschild BH. Beams propagate along the geodesics (red dashed lines).}
    \label{fig:curved_beam_test}
\end{figure}

This test imitates the typical geometry in astrophysical simulations (like an NS). We set the sphere of radius $R=1$ where emissivity and absorption opacities are equal $\eta = \kappa_{\rm A}$. This test is performed in 3D and in Cartesian coordinates $x,y,z \in (-2.5, 2.5)$ and with resolution of $100^3$. Fig.~\ref{fig:rad_sphere_test} shows the radiation energy density $E = n_{\mu} n_{\nu} R^{\mu}_{\nu}$ in steady state (circles) compared with the analytic solution (solid) for two different values of $\eta=\kappa_{\rm A}$. The analytic solution is obtained from integrating the distribution function (where $\mu = \cos\theta$)

\begin{equation}
    \begin{split}
        f (r, \mu) &= \eta \left[ 1 - \exp \left( -\kappa_{\rm A} R s (r, \mu) \right) \right] \\
        s (r, \mu) &= 
            \begin{cases}
            r\mu / R + g(r, \mu), \quad r < R \text{ and } -1 \leq \mu \leq 1, \\
            2 g(r, \mu), \phantom{00000} \quad r \geq R \text{ and } \sqrt{1-(R/r)^2} \leq \mu \leq 1, \\
            0, \phantom{000000000000} \text{otherwise}
            \end{cases} \\
        g (r, \mu) &= \sqrt{1-(1-\mu^2)(r/R)^2}
    \end{split}
\end{equation}

We find a good agreement with analytic solution. This demonstrates that our scheme performs pretty well in case of diverging beams (outside the sphere).

\subsection{Single beam propagation in the curved spacetime}

As we solve the neutrino-transport equations in the vicinity of the BH, we also set up a test to check the propagation of the single beam of neutrinos in curved spacetime. We set the spin parameter to $a_{\rm BH} = 0$ and set up a 2D grid in $r \in (1, 20)\rg, \varphi \in (0, \pi/2)$ at a resolution $150^2$. Next we shine beams of neutrinos emanating from the boundary at $\varphi=0$ with $\gamma_{R} = 250$, such that $\sqrt{-g}R^i_t = 0$ for $i=r,\theta$ and directed along $\hat{\varphi}$ only. We set the beams at different distances from the BH: $r \in (4.8,5.8)\rg$, $r \in (8.6, 9.6)\rg$ and $r \in (16.2, 17.1)\rg$. The results are displayed in Fig.~\ref{fig:curved_beam_test}. We see that beams follow geodesics (red dashed lines) pretty well, and their energy densities undergo some dissipation and widening as they propagate.

\section{Neutrino field evolution}

\begin{figure}[hbtp]
	\centering
	\includegraphics[width=0.95\columnwidth]{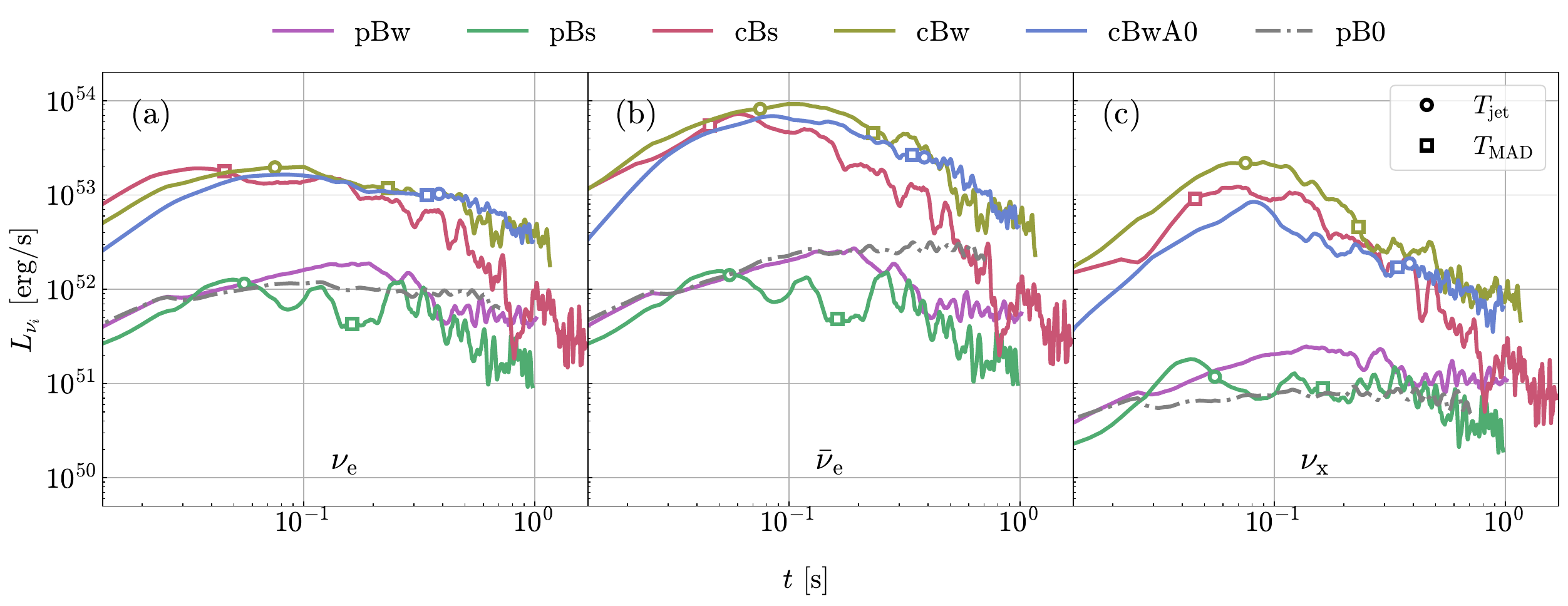}
	\caption{Time evolution of electron neutrino [panel (a)], antineutrino [panel (b)] and heavy lepton neutrino [panel (c)] luminosities for all models. Neutrino luminosities are calculated as radiative power $L_{\nu_i} = \int \sqrt{-g} (-R^r_t) d\theta d\varphi |_{r=r_{\rm ext}}$, measured at an extraction radius $r_{\rm ext} = 100 \rg \approx 600 \rm{km}$. All quantities are smoothed over $\Delta t=10^3 \rg/c=0.02$~s.}
	\label{fig:nu_vs_time}
\end{figure}

\begin{figure}[hbtp]
    \centering
    \includegraphics[width=\textwidth]{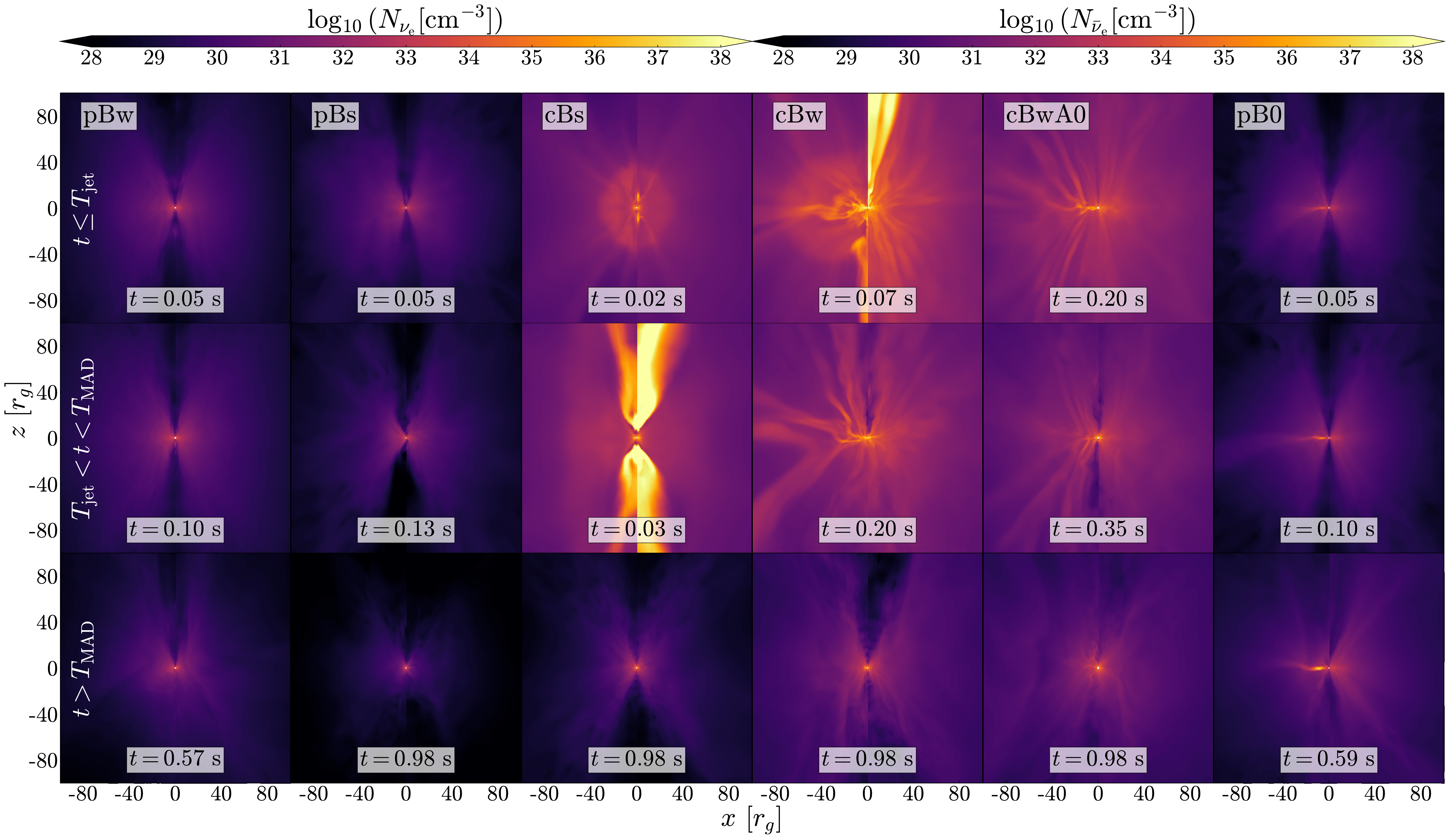}
    \caption{Meridional cuts of the electron neutrino  (left-hand side) and antineutrino number density (right-hand side) for all models (different columns). The top row displays the system before jet launching ($ t < t_{\rm jet} $), middle row -- after the launching of jets and disk winds ($ t_{\rm jet} < t < t_{\rm MAD} $), and the bottom row shows the systems after they achieve a MAD state ($ t > t_{\rm MAD} $).}
    \label{fig:megafigure_nu}
\end{figure}

Figure~\ref{fig:nu_vs_time} shows the electron neutrino, antineutrino and heavy lepton neutrino luminosities measured outside the accretion disk, at an extraction radius of $r_{\rm ext} = 100\,\rg \approx600\,\rm{km}$. Figure~\ref{fig:megafigure_nu} shows the meridional maps of electron neutrino and antineutrino number densities at different times, similar to Figure~\ref{fig:megafigure}.

\newpage
\section{Timescales}

\begin{figure}[hbtp]
	\centering
	\includegraphics[height=3in]{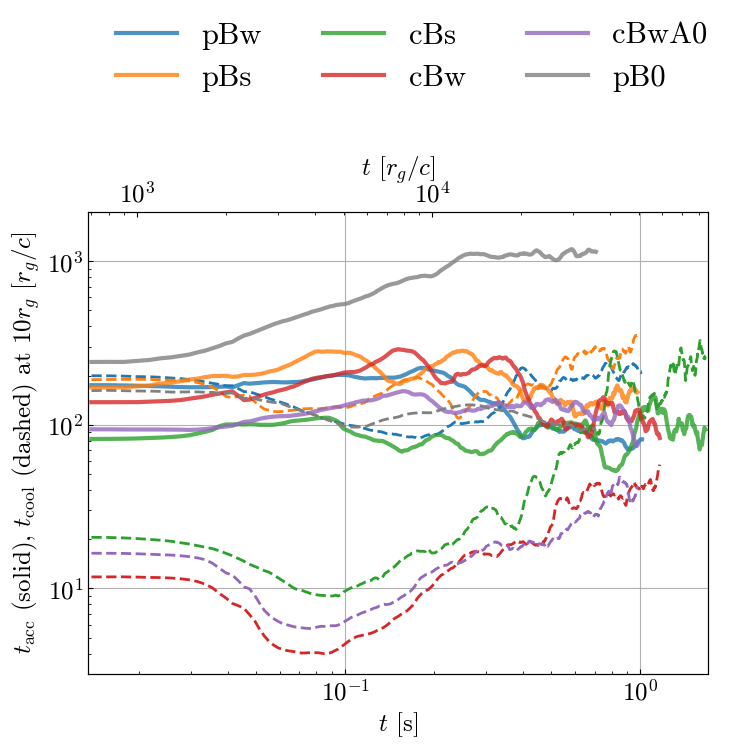}
	\caption{Density-weighted average accretion (solid lines) and neutrino cooling (dashed lines) timescales in the disk at a characteristic radius, $r=10\rg$, in all models, as functions of time.}
	\label{fig:timescales_disk}
\end{figure}

Figure~\ref{fig:timescales_disk} shows the density-weighted accretion and neutrino-cooling timescales in the disk. Longer cooling timescales reflect decreasing disk densities. Together with the shorter accretion timescales once the MAD state is reached, these effects lead to a decrease in the ignition radius.

\end{document}